\documentclass[structabstract]{aa} 
\usepackage[utf8]{inputenc}
\usepackage{txfonts} 
\usepackage{graphicx} 
\usepackage{natbib} 
\usepackage{url} 

\usepackage{graphicx}	
\usepackage{amsmath}	
\usepackage{amssymb}	
\usepackage[dvipsnames]{xcolor}

\usepackage{comment}

\usepackage[colorlinks = true,
            linkcolor = blue,
            urlcolor  = blue,
            citecolor = blue,
            anchorcolor = blue]{hyperref}

\usepackage{ulem}

\usepackage{longtable}

\usepackage{threeparttable}

\usepackage{colortbl}

\usepackage{subcaption}
\usepackage{mwe}

\usepackage{enumitem}

\usepackage{pifont}
\def\msun{\, \mathrm{M}_{\hbox{$\odot$}}}
\def\rsun{\, \mathrm{R}_{\hbox{$\odot$}}}

\newcommand{\be}{\begin{equation}}
\newcommand{\ee}{\end{equation}}

\newcommand{\mstar}{M_{\star}}
\newcommand{\rstar}{R_{\star}}

\newcommand{\rp}{R_{\rm p}}
\newcommand{\rt}{R_{\rm t}}
\newcommand{\rmax}{R_{\rm max}}

\newcommand{\rint}{R_{\rm int}}
\newcommand{\rph}{R_{\rm ph}}
\newcommand{\dsc}{d_{\rm sc\_ph}}

\newcommand{\delth}{\Delta\theta}
\newcommand{\delphi}{\Delta\phi}

\newcommand{\tmin}{t_{\rm min}}

\newcommand{\mdotp}{\dot{M}_{\rm p}}
\newcommand{\mdotfb}{\dot{M}_{\rm fb}}
\newcommand{\mdotout}{\dot{M}_{\rm out}}
\newcommand{\menv}{M_{\rm env}}

\newcommand{\mh}{M_{\rm BH}}

\newcommand{\emin}{e_{\rm min}}

\newcommand{\vout}{v_{\rm out}}

\newcommand{\rints}{R_{\rm int}=10^{14}\, {\rm cm}}
\newcommand{\rintm}{R_{\rm int}=2.5\times10^{14}\, {\rm cm}}
\newcommand{\rintl}{R_{\rm int}=5\times10^{14}\, {\rm cm}}

\newcommand{\rphs}{R_{\rm ph}=10^{14}\, {\rm cm}}
\newcommand{\rphm}{R_{\rm ph}=2.5\times10^{14}\, {\rm cm}}
\newcommand{\rphl}{R_{\rm ph}=5\times10^{14}\, {\rm cm}}

\newcommand{\delths}{\Delta\theta = 45^{\circ}}
\newcommand{\delthm}{\Delta\theta = 90^{\circ}}
\newcommand{\delthl}{\Delta\theta = 180^{\circ}}

\begin{document}

   \title{Modeling continuum polarization levels of tidal disruption events based on the collision-induced outflow model }

    \titlerunning{Modeling continuum polarization levels of TDEs based on the CIO model}

   \author{P. Charalampopoulos\inst{1}\fnmsep\thanks{Contact e-mail: \href{mailto:pngchr@space.dtu.dk}{pngchr@space.dtu.dk}}\href{https://orcid.org/0000-0002-7706-5668}\
          \and
          M. Bulla\inst{2,3,4}\href{https://orcid.org/0000-0002-8255-5127}
          \and
          C. Bonnerot\inst{5}\href{https://orcid.org/0000-0001-9970-2843}
          \and
          G. Leloudas\inst{1}\href{https://orcid.org/0000-0002-8597-0756} \    
          }

   \institute{DTU Space, National Space Institute, Technical University of Denmark, Elektrovej 327, DK-2800 Kgs. Lyngby, Denmark
         \and
         The Oskar Klein Centre, Department of Astronomy, Stockholm University, AlbaNova, SE-10691 Stockholm, Sweden
         \and
         Department of Physics and Earth Science, University of Ferrara, via Saragat 1, I-44122 Ferrara, Italy
         \and
         INFN $-$ Sezione di Ferrara, via Saragat 1, I-44122 Ferrara, Italy
         \and
         Niels Bohr International Academy, Niels Bohr Institute, Blegdamsvej 17, DK-2100 Copenhagen Ø, Denmark
             }

   \date{Received - ; accepted -}

 


  \abstract
   {Tidal disruption events (TDEs) have been observed in the optical and ultraviolet (UV) for more than a decade but the underlying emission mechanism still remains a puzzle. It has been suggested that viewing angle effects could potentially explain their large photometric and spectroscopic diversity. Polarization is indeed sensitive to the viewing angle and the first polarimetry studies of TDEs are now available, calling for a theoretical interpretation. In this study, we model the continuum polarization levels of TDEs using the 3-D Monte Carlo radiative transfer code \textsc{possis} and the collision-induced outflow (CIO) TDE emission scenario where unbound shocked gas originating from a debris stream intersection point offset from the black hole, reprocesses the hard emission from the accretion flow into UV and optical bands. We explore two different cases of peak mass fallback rates $\mdotp$ ($\sim3 \,\msun\,\rm yr^{-1}$ and $\sim0.3 \,\msun\, \rm yr^{-1}$) while varying the following geometrical parameters: the distance $\rint$ from the black hole (BH) to the intersection point (where the stellar debris stream self intersects), the radius of the photosphere around the BH $\rph$, on the surface of which the optical/UV photons are generated, and the opening angle $\delth$ that defines the fraction of the surface of the photosphere that photons are generated on (anisotropic emission). For the high mass fallback rate case, we find for every viewing angle polarization levels below one ($P<1\%$) and $P<0.5\%$ for 10/12 simulations. The absolute value of polarization reaches its maximum ($P_{\rm max}$) for equatorial viewing angles. For the low mass fallback rate case, the model can produce a wide range of polarization levels for different viewing angles and configurations. The maximum value predicted is $P\approx8.8\%$ and $P_{\rm max}$ is reached for intermediate viewing angles. We find that the polarization depends strongly on i) the optical depths at the central regions (between the emitting photosphere and the intersection point) set by the different $\mdotp$ values and ii) the viewing angle. With time, there is a drop in densities and optical depths leading to general increase in polarization values and $P_{\rm max}$, although the opposite trend can be observed for specific viewing angles. Increasing the distance $\rint$ between the intersection point and the black hole seems to generally favour higher polarization levels. Finally, by comparing our model predictions to polarization observations of a few TDEs, we attempt to constrain their observed viewing angles and we show that multi-epoch polarimetric observations can become a key factor in constraining the viewing angle of TDEs.}
   

   \maketitle
%

\section{Introduction} \label{sec:intro}

A tidal disruption event (TDE) happens when the tidal radius (R$_{\rm t}$) of a supermassive black hole (SMBH) 
intersects the trajectory of an orbiting star whose pericenter distance (R$_{\rm p}$) is smaller than R$_{\rm t}$ where R$_{\rm t}$~$\approx$~R$_{\rm *}$(M$_{\rm BH}$/M$_{\rm *}$)$^{1/3}$ and R$_{\rm *}$ and M$_{\rm *}$ are the stellar radius and mass and M$_{\rm BH}$ is the mass of the SMBH \citep{Hills1975}. The immense gravitational field of the black hole leads to a large spread in the specific orbital binding energy of the star (which is greater than its mean binding energy) and the star gets ripped apart in a TDE \citep{Rees1988}. The stellar debris are stretched into a thin elongated stream and around half of it stays bound to the SMBH and falls back towards it on highly eccentric orbits \citep{Rees1988,Evans1989}. As the debris circularize around the black hole, a strong, transient flare is produced \citep{Lacy1982,Rees1988,Evans1989,Phinney1989} with L$_{\rm bol} \sim$ 10$^{41-45}$ erg~s$^{-1}$, which sometimes emits above the Eddington luminosity \citep{Strubbe2009,Lodato2011}. Even though the occurrence of TDEs was predicted by theorists almost five decades ago \citep{Hills1975}, observations of such exotic transients happened much later, first in the X-ray regime \citep{Komossa1999}, then in the ultraviolet (UV) \citep{Gezari2006}, and finally in the optical wavelengths \citep{Gezari2012}. Furthermore, there are TDEs discovered in the mid-infrared \citep{Mattila2018,Kool2020,Jiang2021,Reynolds2022} and others that launch relativistic jets and outflows leading to bright gamma/X-ray and radio emission (e.g., \citealt{Zauderer2011,VanVelzen2016,Alexander,Goodwin2022}). 
When the bound debris passes at pericenter, relativistic precession causes a self-intersection of the debris stream and energy gets dissipated \citep{Strubbe2009,Shiokawa2015,Guillochon2015,Bonnerot2020}. Since TDEs were considered to be accretion-powered events \citep{Komossa2002}, they were expected to peak at the X-ray wavelengths. However, there seems to be a large diversity in the X-ray properties of TDEs; no X-ray emission has been detected from up to 50\% of TDE candidates, some others emit primarily in the X-rays and some ``intermediate cases'' show both moderate X-ray alongside with optical/UV emission with X-ray to optical ratios spanning the entire range between $\leq$ 10$^{-4}$ to $\geq$ 10$^{3}$ \citep{Auchettl2017}. This large diversity remains to be fully explained through an understanding of the underlying emission mechanism of those transient flares.

Two main families of models have been proposed to explain such strong optical/UV emission (luminosities $\sim 10^{44}$ erg~s$^{-1}$) without accompanying X-rays. The first scenario suggests that the accretion disk emission is reprocessed to less energetic wavelengths by material around the SMBH (e.g., \citealt{Loeb1997,Strubbe2009,Guillochon2014,Roth2016}). A unification scenario of TDEs has been proposed \citep{Dai2018,Thomsen2022} which describes their geometry with a thick, super-Eddington accretion disk. Because of inefficient accretion (mostly at early times) a polar relativistic jet as well as optically thick outflows of material are launched \citep{Metzger2016} and thus reprocess radiation to longer wavelengths. Depending on the line of sight of the observer, a TDE can be perceived as ``optical'' if viewed edge-on (all X-rays are reprocessed) or as ``relativistic/X-ray'' if viewed face-on (X-rays escape from the outflow/jet/funnel). Intermediate angles can reveal both optical and X-ray emission and it has been suggested that their spectral properties are also viewing-angle dependent \citep{Charalampopoulos2022}. A second scenario proposes that the optical/UV emission is produced by the shocks arising from the debris streams collision/self-intersection \citep{Piran2015,Jiang2016}. In this scenario, fluctuations in the intersection point drive material to the center (the stream collision occurs off-center) later to form an accretion disk and emit the (sometimes) observed X-rays (which are delayed compared to the optical/UV emission \citealt{Pasham2017}). In both scenarios the bound debris are forming a photosphere -- either surrounding the SMBH or the intersection point -- and outflows can be launched -- either because of inefficient accretion or because unbound material leaves the self-intersection point where the stream collides.

 \citet{Lu2020} show that unbound debris can be created from the shock occurring at the self-crossing debris stream due to relativistic precession as it returns the pericenter. In their model, termed ``collision-induced outflow'' (CIO), the optical/UV radiation does not arise because of the shock produced at the self-intersecting point of the debris stream but from accretion of infalling matter from the intersection point towards the BH. The returning stream of debris intersects itself, causing the formation of a quasi-spherical outflow that determines the density distribution at large radii. Radiation produced by accretion near the black hole has to diffuse through this surrounding matter before it emerges from the system.The EUV/X-ray radiation from the accretion is reprocessed by the CIO and re-emitted to optical/UV wavelengths. 
 Using radiation-hydrodynamics simulations, \citet{Bonnerot2021} showed that when this envelope (which is offset with respect to the black hole) gets irradiated by the photons produced by shocks near the black hole, there is always a radiation-free region remaining around the intersection point (see their Figure 3). In this region, the gas will have a low ionization level resulting in increased bound-free opacities. They suggest that photons diffusing towards this direction therefore have a higher likelihood (compared to the case of an envelope centered on the black hole) of being absorbed and then re-emitted in the optical band before emerging from the system.

One way to test the different competing models for TDEs is through their polarimetric properties, a method that has been widely used in the past to probe the physics and geometry of transients. 
The source and degree of polarization vary depending on the transient and the underlying physical mechanism. 
In supernovae (SNe), polarization is mainly caused by electron scattering and the continuum polarization is determined by how asymmetrical the photosphere is \citep{Hoflich1991,Kasen2003,Wang2008,Patat2017}.
Excluding relativistic TDEs, where the source of polarization has been attributed to synchrotron radiation \citep{Wiersema2012,Wiersema2020}, 
until recently there had been very few polarimetric studies on optical TDEs, where the data was either very sparse 
\citep{Higgins2019,Lee2020} or focused on peculiar events \citep{Maund2020}.

The study of \citet{Leloudas2022} presented spectral polarimetry for a sample of 3 optical TDEs. After carefully correcting for the effect of light dilution from the host galaxy, AT~2018dyb was measured to have a continuum polarization of 2.1\% at $-17$ days (with respect to peak), dropping to 1.2--1.3\% at $+40$ and $+50$ days.
The polarization of AT~2019dsg decreases rapidly from $(6.1 \pm 1.4)$\% at $+$16 days to $\sim 1.0 - 1.2$\% at $+31 - 74$ days. AT~2019azh has a lower continuum polarization of $0.7$\% in a single epoch (+22d). 
All these results have been carefully corrected for light dilution from the TDE host galaxy and for the ISP. 
In another simultaneous study \citet{Patra2022} found the polarization of AT~2019qiz to increase from $\sim$ $(0.16 \pm 0.15)$\% at peak to $\sim$ ($0.93 \pm 0.19$)\% at $+29$ days.
These values have not been corrected for the host galaxy contamination but they indicate the opposite trend (i.e. an increase in polarization) than the TDEs in \citet{Leloudas2022}. Recently (while this work was in a very late stage) another imaging polarimetry TDE study appeared in the literature, of TDE AT~2020mot \citep{Liodakis2022}. The authors (after correcting for the host galaxy contamination) measure the striking value of $P=(25\pm4)\%$ for a single epoch and suggest that the optical luminosity of AT~2020mot is powered by shocks during the tidal stream collisions, since they claim that such high optical polarization cannot be achieved by reprocessing.

Furthermore, \citet{Leloudas2022} demonstrate that the continuum polarization is wavelength independent, and that the spectrum is depolarized by emission lines while polarization peaks can be present at the broad line wings. Consequently they suggest that electron scattering is the primary cause of polarization in optical TDEs and rule out synchrotron radiation and dust scattering as significant factors. The late-time data can be approximately fitted with a dominant axis, indicating that the TDEs settle to an axisymmetric shape/geometry. The authors suggest that this is compatible with optically thick outflows and the formation of an accretion disk and they model the polarization with the super-Eddington accretion model of \citet{Dai2018} and the radiative transfer code \textsc{possis} \citep{Bulla2019a}. Their modeling finds that extended disks provide polarization predictions that are broadly consistent with the observations and that the polarization
signal primarily depends on the total mass included in the disk, the compactness of the disk (density/optical depths), and the viewing angle.
 
Naturally, we want to test whether other emission mechanism scenarios are compatible with the data (and the future data) of the growing field of TDE (spectro)polarimetry and whether we can reject those models or use them in order to constrain the vast parameter space. In this paper, we present the continuum polarization modeling of the CIO scenario. 

We introduce our model in Sect. \ref{sec:model} while describe the setup of our radiative transfer calculations in Sect. \ref{sec:rts}. We summarize our results in Sect. \ref{sec:results} and we discuss their implications in Sect. \ref{sec:discussion}. Sect. \ref{sec:conclusion} contains our summary and conclusions. 


\section{Model} \label{sec:model}

\begin{figure*}
\centering
\includegraphics[width=0.8 \textwidth]{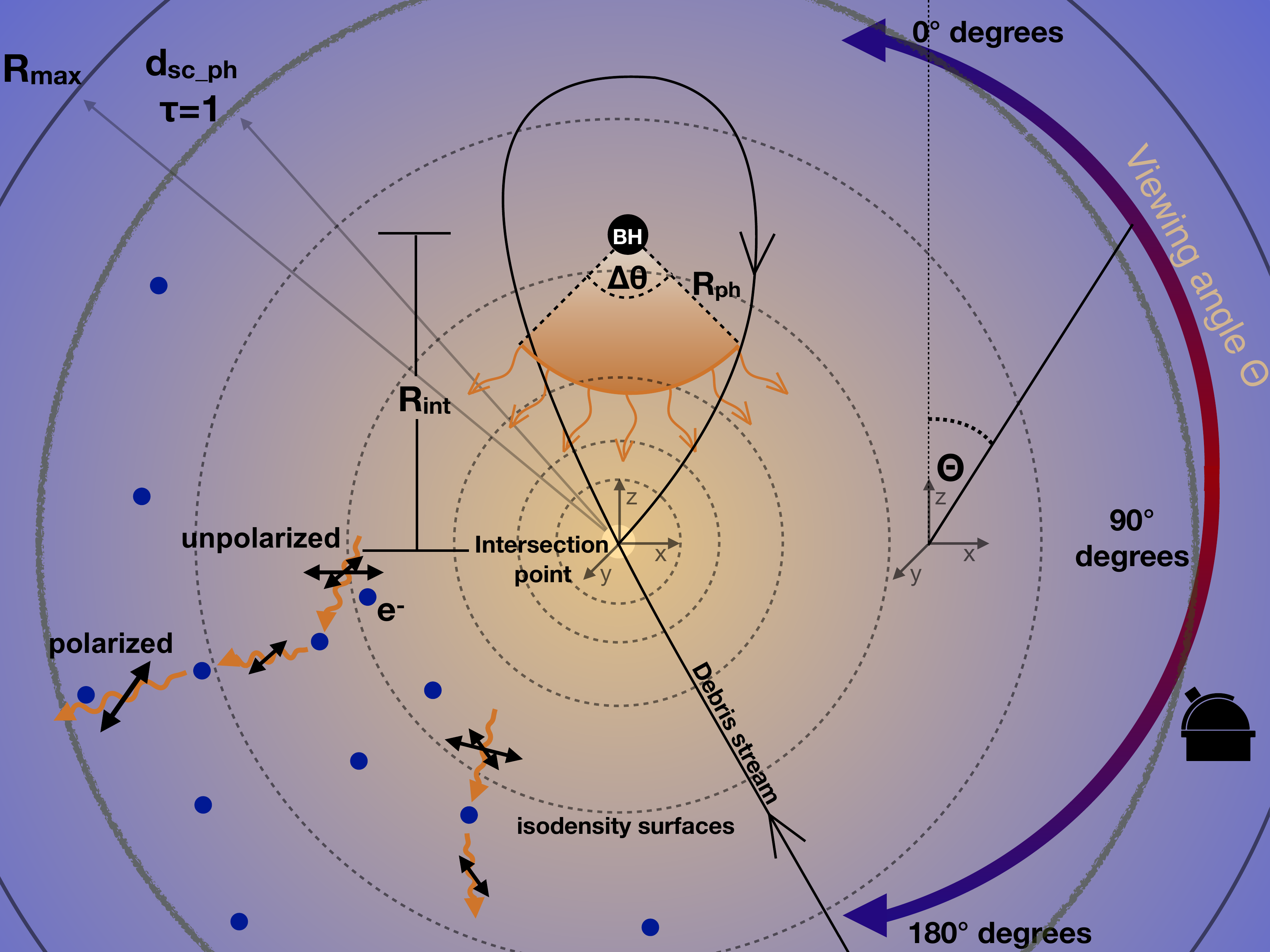}
\caption{A schematic illustration (not to scale) of the CIO scenario \citep{Lu2020} modeled in this work. The black hole lies at a distance $\rint$ from the debris stream intersection point from which the CIO is launched. Photons are generated on a sphere of radius $\rph$ centered on the black hole and within a cone of opening angle $\delth$ aligned with the negative z axis. The electron scattering photosphere where the optical depth is $\tau=1$ is located at a distance $\dsc$ from the intersection point while the surrounding sphere of radius $\rmax$ corresponds to the outer radius of our grid. We exploit the model symmetry about the $z$ axis and extract polarization levels for 21 viewing angles in the $xz$ plane, equally spaced in cosine ($\Delta\cos\Theta=0.1$) between a north-polar ($\cos\Theta=1$, face-on) and a south-polar ($\cos\Theta=-1$, face-off) orientation. Photons are unpolarized when they are injected from the photosphere and they become polarized after scattering with an electron (visualized as blue circles). Depending on the direction they travel, they are scattered towards the observer either with an electric field oscillating in the horizontal direction (i.e. with a negative $q$, see Equation \eqref{eq:stokesdef}) or with an electric field oscillating in the vertical direction (i.e. with a positive $q$).}
\label{fig:cartoon}
\end{figure*}

The TDE emission mechanism that we model in this work is the CIO model \citep{Lu2020,Bonnerot2021}. The overall idea is that the returning stream of debris intersects itself, causing the formation of a quasi-spherical outflow that determines the density distribution at large radii. Here we will provide an analytical description for the density distribution in tidal disruption events (TDEs) close to the moment they reach their luminosity peak as presented in \citet{Bonnerot2021}.

 The pericenter distance of the star which is initially on a parabolic trajectory, is given by ${\rp=\rt/\beta}$, where $\beta \geq 1$ is the penetration factor that specifies the depth of the encounter inside the tidal radius given by
\be
\rt = \rstar \left(\frac{\mh}{\mstar}\right)^{1/3},
\label{eq:tidal-radius}
\ee
During the disruption, the stellar debris undergo a spread in orbital energy of $\Delta \epsilon = G \mh \rstar / \rp^2$ \citep{Stone2013} and therefore evolve into an elongated stream, half of which is bound to the black hole and therefore falls back to it after reaching the apocenter. The rate at which this bound debris returns to pericenter is
\be
\mdotfb = \mdotp \left(1+\frac{t}{\tmin}\right)^{-5/3},
\label{eq:mdotfb}
\ee
which peaks at a value of 
\be
\mdotp = 3\msun \rm yr^{-1} \left(\frac{\mh}{10^{6}\msun}\right)^{-1/2} \left(\frac{\mstar}{\msun}\right)^{2} \left(\frac{\rstar}{\rsun}\right)^{-3/2} \equiv \frac{\mstar}{3 \tmin},
\label{eq:calcs}
\ee

where 
\be
\tmin = 2 \pi G \mh (2 \Delta\epsilon)^{-3/2} = 2^{-1/2} \pi \left(\frac{G\mstar}{\rstar^{3}}\right)^{-1/2} \left(\frac{\mh}{\mstar}\right)^{1/2} 
\label{eq:tmin}
\ee
denotes the orbital period of the most bound debris \citep{Lodato2011}, the first to fall back to the black hole with an eccentricity $\emin$ given by
\be
1 - \emin =  \frac{2}{\beta} \left(\frac{\mh}{\mstar}\right)^{-1/3}.
\ee
As it passes at the pericenter, the tip of the returning stream experiences relativistic apsidal precession which makes its orbit precess by an angle
\be
\delphi \approx \frac{3\pi R_{g}}{\rp}
\ee
where $R_{g} = G \mh/c^{2}$ is the gravitational radius. As a result, this part of the stream which moves away from the black hole intersects with the still approaching matter, at an intersection radius of \citep{Dai2015}
\be
\rint =  \frac{\rp(1+\emin)}{1-\emin\cos{(\delphi/2)}},
\label{eq:rint}
\ee
that decreases from apocenter to pericenter for an increasing precession angle. This interaction leads in the so-called self-crossing shock that launches the matter into a quasi-spherical outflow centered on the intersection point. We can approximate the outflow velocity by the local escape speed which is given by
\be
\vout =  \left(\frac{2G\mh}{\rint}\right)^{1/2}.
\ee
This constant velocity, used for simplicity here, does not take into account the effect of the gravitational force from the black hole and imposes that outflowing matter is unbound while more detailed calculations \citep{Lu2020} find that part of it stays bound. Throughout this work, we will consider a constant outflow velocity of ${\vout=0.03c}$.

A realistic mass rate profile for this outflow would be a function of both distance $r$ from the intersection point, and time $t$ which we call $\mdotout(r,t)$. The value of this function at $r$ = 0 is given by the fallback rate, i.e. $\mdotout(0,t)$ = $\mdotfb(t)$, while at $r\neq$ 0, we have $\mdotout(r,t)$ = $\mdotout(0,t-r/\vout))$. This last equality comes from mass conservation applied to a fluid element launched from the intersection point at a time $t' = \rm t-r/\vout$. Combining these relations leads to 
\be
\mdotout(r,t) = \mdotfb(t-r/\vout). 
\label{eq:mdotout}
\ee
Equation \eqref{eq:mdotfb} (used to compute the fallback rate), does not account for what happens during the rise to the peak value (i.e. when $\rm t<0$) as it assumes that the fallback rate only decreases (we consider $\mdotfb(t)=0$ if $t<0$). Hence we do not model the rise to the peak. The fact that the fallback rate is only non-zero for $t>0$ translates to $t' = t-r/\vout > 0$ hence $r<t\cdot \vout$. This condition makes the outflow rate $\mdotout$ only non-zero below a maximum radius $\rmax=t\cdot \vout$, which will be the outer radius of of the outflow we consider. 

The density of this outflowing gas can be described as
\be
\rho_{\rm out} =  \frac{\mdotout}{4\pi\vout r^{2}} = \frac{\mdotp}{4\pi\vout r^{2}}\left[1+\frac{(\vout\cdot t) - r}{\vout\cdot \tmin}\right]^{-5/3}
\label{eq:density_time}
\ee
where $r$ is the distance from the intersection point whose exact location is given by Equation \eqref{eq:rint}. Therefore the density is a function of distance and time. 
We do not consider the presence of an accretion disc that should form around the black hole as outflowing gas is reaching it \citep{Bonnerot2020}, which is legitimate since it probably only affects the density at very close distances from the black hole. The density distribution given by Equation \eqref{eq:density_time} constitutes a first-order description which has the advantage of being entirely analytic. 

We suppose that radiation produced on the surface of a photosphere of size $\rph$ around the black hole, has to diffuse through this outflowing gas before emerging from the system. In order to simulate the absorption and re-emission of X-ray photons in the radiation-free, low-ionization, high density region around the intersection point, we suppose that optical photons originate from only a fraction $\delth$ of the surface of the emitting photosphere $\rph$. In this way, we model the reprocessing of X-rays to optical in this region \citep{Bonnerot2021} by directly injecting optical photons towards this direction, radially away from the black hole.

Throughout this work we will consider the following two representative TDE cases:
\be
  \mdotp=\begin{cases}
    2.93 \,\msun\, \rm yr^{-1} \,\,(\rm{Case \,A})\\
    0.29 \,\msun\, \rm yr^{-1} \,\,(\rm{Case \,B})
  \end{cases}
\ee
which are reasonable peak fallback rate values for TDEs, Case A at the higher and Case B at the lower end (see \citealt{Law-Smith2020}). Those $\mdotp$ values can be reproduced by substituting various combinations of $\mh$, $\mstar$ and $\rstar$ into Equation \eqref{eq:calcs}. For example, two indicative cases would be, one with $\mh=1.5\times10^{6}\msun$ and $\mstar=1\msun$ (Case A) and one with $\mh=5\times10^{6}\msun$ and $\mstar=0.1\msun$ (Case B). Then, we can determine the $\rstar$ of those two cases as function of $\mstar$ by using the mass–radius relation for ZAMS solar-metallicity stars given in \citet{Tout1996} for main-sequence stars, which result in $\rstar=0.88\rsun$ and $\rstar=0.12\rsun$, respectively. If we substitute those values into Equation \eqref{eq:tmin}, we get values of $\tmin\approx40.55$ days for Case A and $\tmin\approx41.67$ days for Case B. In order to set up our grid, we use Equation \eqref{eq:density_time} and we consider $t=\tmin$ which would be a value close to the peak of a TDE light curve.

While electron scattering polarizes radiation, bound-free and free-free transitions can depolarize the radiation and potentially introduce a wavelength dependence in the overall polarization signal. In this work we neglect the wavelength-dependent free-free and bound-free opacities and we study a pure electron scattering case. In practice, this means that we neglect thermalization effects and assume full-ionization. This assumption is the same as in \citet{Leloudas2022} and is motivated by their findings that show a lack of a significant wavelength dependence in the continuum polarization of TDEs. Furthermore, full-ionization in the reprocessing photospheres of TDEs has been predicted by various theoretical works (see e.g. \citealt{Metzger2016,Roth2016,Dai2018}). \citet{Metzger2016} show that the neutral fraction in the envelope of TDEs is very small (<$10^{-10}$) because the gas is strongly ionized by the central radiation flux. Such small neutral fraction leads to bound-free opacities smaller than $10^{-4}$\,cm$^2$\,g$^{-1}$. Assuming solar composition (hydrogen mass fraction $X=0.7$) as in previous works, 
we set the electron-scattering opacity to $\kappa_\mathrm{es}= 0.2 \times(1+X)=0.34$\,cm$^2$\,g$^{-1}$. This implies a small ratio of the bound-free to electron scattering opacity, hence justifying our full-ionization assumption. Following a simple calculation (integrating radially from infinity inwards) the location of the scattering photosphere can be found from:
\be
1 = \int_{\dsc}^{\infty} \rho\kappa_{es} \,dr 
\ee
The outer radius of our grid $\rmax$ is always located at ${\rmax=t \cdot \vout}$ and for some cases, it is located within the scattering photosphere or to phrase it in a different way; the density in the outflow is large enough that $\tau>1$ everywhere inside $\rmax$. The two mass outflow rate values that we test in this work are translated to envelope masses $\menv$ of:\be
  \menv=\begin{cases}
    0.271\,\msun \,\,(\rm{Case \,A})\\
    0.026\,\msun \,\,(\rm{Case \,B})
  \end{cases}
  \label{eq:menvs}
\ee
for $t=\tmin$.

We present a schematic visualisation of our model and its parameters in Fig. \ref{fig:cartoon}.

\section{Radiative transfer and polarization simulations} \label{sec:rts}
To predict the polarization levels, we use the 3-D Monte Carlo radiative transfer code \textsc{possis} \citep{Bulla2019a}. This code has been used in the past to predict polarization for astrophysical transients such as supernovae \citep{Bulla2015,Inserra2016}, kilonovae \citep{Bulla2019,Bulla2021} and TDEs \citep{Leloudas2022}. The code accommodates arbitrary 3-D geometries and simulates the propagation of $N_{\rm ph}$ Monte Carlo photon packets as they diffuse outward through the expanding material. Each packet is assigned a Stokes vector 
\begin{equation}
\label{eq:stokesdef}
\textbf{S} = \begin{pmatrix} I \\ Q \\ U \end{pmatrix} = \begin{pmatrix}  \updownarrow + \leftrightarrow \\ \updownarrow - \leftrightarrow  \\ \mathrel{\rotatebox{45}{$\updownarrow$}}-\mathrel{\rotatebox{45}{$\leftrightarrow$}}  \end{pmatrix} ,
\end{equation}
where $I$ is the total intensity and $Q$ and $U$ measure the linear polarization as the difference in intensities between two orthogonal directions.
The normalized Stokes vector ${\bf s}=(1,q,u)$ is initialised to ${\bf s_0}=(1,0,0)$ and updated after every interaction with matter. Here, we perform simulations with (Thomson) electron scattering as the only source of opacity 
as mentioned in Sect.~\ref{sec:model}. Polarization levels as a function of viewing angle are extracted using the ``virtual-packet'' approach introduced by \cite{Bulla2015} as this is a more efficient technique compared to the angular binning of escaping photons typically adopted in standard Monte Carlo simulations. 

\begin{table}
\renewcommand{\arraystretch}{1.2}
\setlength\tabcolsep{0.05cm}
\fontsize{9}{11}\selectfont
\caption{Details of each simulation case presented in this work for three different time snapshots. $\mdotp$ is the peak fallback rate, $\menv$ is the integrated envelope mass which increases with time, $\rmax$ is the outer radius of the outflow/grid, $\dsc$ is the distance of the electron scattering photosphere from the center of the grid (the intersection point), $\Delta \rm r$ is the grid resolution and $N_{\rm cells}$ is the number of cells in the grid.
}\label{tab:sims}
\begin{tabular}{c | c c c c c c}
\hline
 & $\mdotp$  & $\menv$ & $\rmax$ & $\dsc$ & $\Delta r$ & $N_{\rm cells}$ \\
 & ($\,\msun\, \rm yr^{-1}$)& ($\msun$) &  ($10^{15}$ cm)&  ($10^{15}$ cm) &  ($10^{13}$ cm) & \\
\hline
$t=\tmin$ &  & &  &  &  &  \\
\hline
Case A & 2.93 & 0.271 & 3.15 & >$\rmax$  & 2.5 & 252 \\
Case B & 0.29 & 0.026 & 3.15 & 0.26  & 2.5 & 252 \\
\hline
$t=\tmin+25\rm d$ & & &   &  &  &  \\
\hline
Case A & 2.93 &  0.346 & 5.09 & >$\rmax$  & 3.5 & 290 \\
Case B & 0.29 & 0.035 & 5.09 &  0.13  & 3.5 & 290 \\
\hline
$t=\tmin+50\rm d$ & & &   &  &  &  \\
\hline
Case A & 2.93 & 0.435 & 7.04 & 3  & 5 & 282 \\
Case B & 0.29 &0.041 & 7.04 &  0.09  & 5 & 282 \\
\hline
\end{tabular}
\\[-0pt]
\end{table}

As shown in Fig.~\ref{fig:cartoon}, the intersection point is placed at the origin of the modeled grid and the black hole at Cartesian coordinates $(x,y,z)=(0,0,\rint)$. Photon packets are generated on a sphere of radius $\rph$ centered on the black hole and within a cone of opening angle $\delth$ aligned with the negative z axis. The photons created on this photospheric surface are assumed to initially propagate radially away from the black hole. For instance, an opening angle $\delthl$ defines emission from the surface of the south hemisphere of the emitting photosphere. We exploit the model symmetry about the $z$ axis and extract polarization levels for 21 viewing angles in the $xz$ plane (i.e. azimuthal angle $\phi=0$), equally spaced in cosine ($\Delta\cos\Theta=0.1$) between a north-polar ($\cos\Theta=1$, face-on) and a south-polar ($\cos\Theta=-1$, face-off) orientation. Because of the axial symmetry, the polarization signal is carried only by the Stoke parameter $q$ while $u$ is taken as a proxy for the Monte Carlo noise in the simulations. $P$ is defined as $P=\sqrt{q^{2}+u^{2}}$, which in the case of $u=0$ simplifies to $P=|q|$. 

Our model contains 3 free parameters; $\rint$, $\rph$ and $\delth$.
In this work we chose to model opening angles of $\delths,\, 90^{\circ}\, \rm{and}\, 180^{\circ}$. For the intersection radius, we modeled reasonable values \citep{Lu2020,Bonnerot2021} of $\rints,\, \rintm\, \rm{and}\, \rintl$ and for the radius of the emitting photosphere values of $\rph=10^{14}\,{\rm cm},\, \rphm\, \rm{and}\, \rphl$. This corresponds to 12 simulations for each of the two mass outflow rate cases (A and B, hence 24 simulations) in order to map all the parameter space. 
In the Appendix, we also include the simulation results for the cases of ${\rph=1\times10^{14}\, {\rm cm}}$, ${\rint=1\times10^{14}\, {\rm cm}}$ and ${\rph=3\times10^{14}\, {\rm cm}}$, ${\rint=3\times10^{14}\, {\rm cm}}$ (for all three $\delth$) so 6 more simulations for each case. For each of the cases we used a different number of injected photons as, the higher the mass, the simulations become more computationally expensive. $N_{\rm ph}=10^{7}$ photons were simulated for the Case A simulations and $N_{\rm ph}=10^{6}$ for the Case B ones. For both Cases we used a grid resolution $\Delta r = 2.5\times10^{13}\, {\rm cm},$ ($N_{\rm cells}=252$). If we wish to study the time evolution of our model and take snapshots at times $t>t_{\rm min}$, the embedded mass $\menv$ of the system rises and we need to take account of that by slightly increasing $\Delta r$ in order to lower the simulation running time. All the above are summarized in Table \ref{tab:sims}.

\section{Results} \label{sec:results}

\subsection{Polarization predictions for the two different mass outflow rate cases} \label{subsec:results}

\begin{figure*}
\centering
\includegraphics[width=0.935 \textwidth]{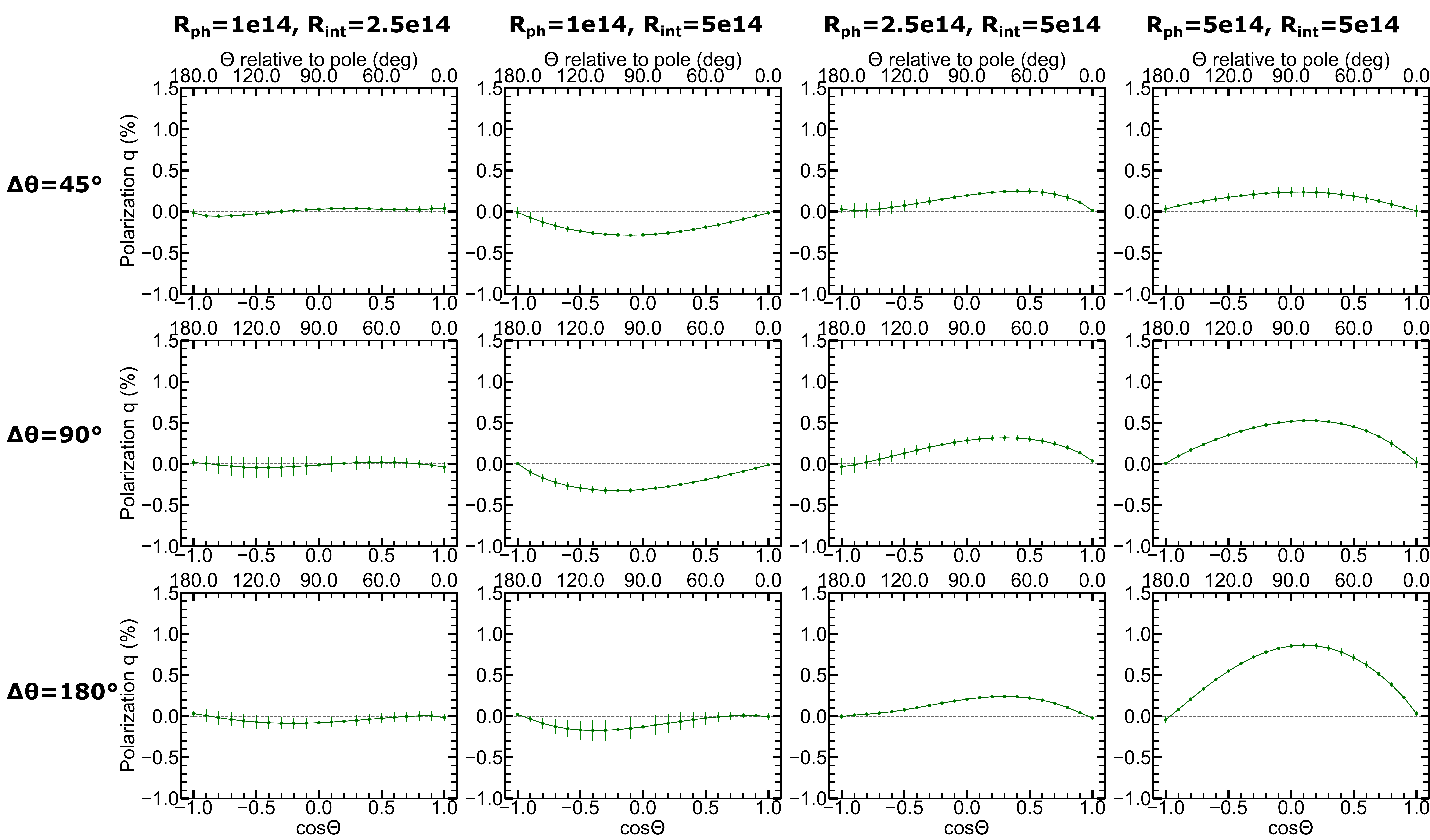}
\caption{Polarization levels for the 12 simulations of the high mass outflow rate scenario (Case A, $\mdotp = 2.93 \,\msun\, \rm yr^{-1}$). The different columns of the plot represent the different combinations of $\rph$ and $\rint$ that we modeled in this work and the different rows represent the three different $\delth$ (see Sect. \ref{sec:model} for an explanation of the choice of those values). Each subplot shows the degree of the polarization $P$ as a function of the 21 viewing angles $\Theta$, equally spaced in cosine ($\Delta\cos\Theta=0.1$) between a north-polar ($\cos\Theta=1$, face-on) and a south-polar ($\cos\Theta=-1$, face-on) orientation. We find that this configuration results in polarization levels below one ($P<1\%$) for all viewing angles and for 10/12 simulations (P<0.5\%).}
\label{fig:med_mass_subs}
\end{figure*}


\begin{figure*}
\centering
\includegraphics[width=0.935 \textwidth]{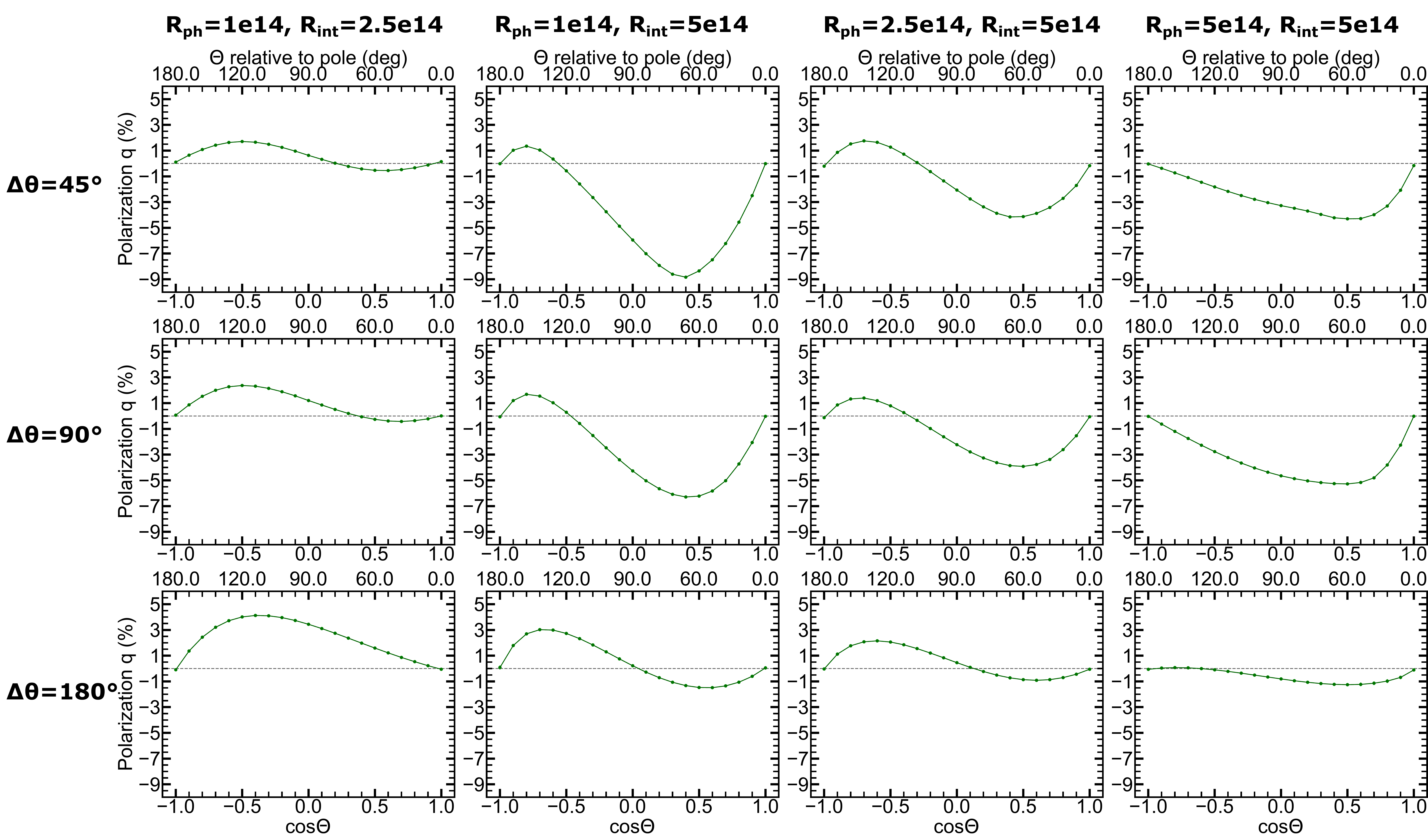}
\caption{Same as Fig. \ref{fig:med_mass_subs} but for the low mass outflow rate scenario (Case B, $\mdotp = 0.29 \,\msun\, \rm yr^{-1}$). We find that this configuration can predict high polarization values (up to $P\approx8.8\%$). The absolute value of polarization reaches its maximum for closer to polar viewing angles ($0.4\leq|\cos\Theta|\leq0.7$ or $45^{\circ}\leq\Theta\leq66^{\circ}\,\&\,114^{\circ}\leq\Theta\leq135^{\circ}$). }
\label{fig:low_mass_subs}
\end{figure*}

The resulting continuum polarization degree $q$ as a function of the viewing angle for every single simulation is presented in Figs. \ref{fig:med_mass_subs} and \ref{fig:low_mass_subs} for Case A and Case B respectively. In all simulations, $q=0$ for both polar orientations ($\cos\Theta=1$ and $\cos\Theta=-1$, face-on and face-off inclinations) due to the axial symmetry of the model, while non-zero polarization signals are found for different orientations. 
Two more combinations of $\rint$ and $\rph$ for Case A (Fig. \ref{fig:med_mass_all_new}) and Case B (Fig. \ref{fig:low_mass_all_new}) are presented in the Appendix. 

For the vast majority of the simulations of Case A, we see the absolute value of polarization to reach its maximum ($P_{\rm max}$) for equatorial viewing angles ($-0.4\leq\cos\Theta\leq0.4$ or $114^{\circ}\leq\Theta\leq66^{\circ}$, where $\theta$ = 90 is the equator) while for Case B, we see the highest values for closer to polar viewing angles ($0.4\leq|\cos\Theta|\leq0.7$ or $45^{\circ}\leq\Theta\leq66^{\circ}\,\&\,114^{\circ}\leq\Theta\leq135^{\circ}$). For the high mass Case A, the CIO model is found to produce polarization below one per cent ($P<1\%$) for every viewing angle. In fact for most simulations (10/12) it is even below 0.5\%. In practice, such values are hard to accurately measure as polarimetry requires very high signal to noise (and therefore bright targets) and the uncertainty in the measurement can be of this order. For the low mass Case B, the model can produce a wide range of polarization levels for different viewing angles and configurations. The maximum value predicted for all the configurations we tested is $P\approx8.8\%$. Four simulations predict a maximum value of $P>4\%$ while all the rest predict maximum values of at least $P>1\%$. 

In general, when photons travel through high optical depth material ($\tau \gg 1$) they undergo multiple scatterings with electrons which causes a loss of information on directionality and thus destroy the polarization signal \citep{Kasen2003}. On the other hand, photons that travel through material of $\tau \sim 1$ are less affected by multiple scatterings (see Sect. \ref{subsubsec:optdsandvas} for further discussion on the effects of varying optical depth as well as indicative optical depths at various distances $r$ from the intersection point in Table \ref{tab:densandtau} in the Appendix). Depending on the direction they travel, they are scattered towards the observer either with an electric field oscillating in the horizontal direction (i.e. with a negative $q$, see Equation \eqref{eq:stokesdef}) or with an electric field oscillating in the vertical direction (i.e. with a positive $q$). For example, for an observer at $\Theta=90^{\circ}$ in the $xz$ plane (see Fig. \ref{fig:cartoon}), the horizontal direction (negative $q$) is defined as photons traveling and scattering along the $z$-axis, while the vertical direction (positive $q$) is defined as photons traveling and scattering along the $y$-axis. If we project the spherical outflow/grid into a plane orthogonal to the viewing angle direction, that would lead into a circle (for any given viewing angle). Now if we divide this circle into four quadrants, an observer looking along a given viewing angle, would see horizontally polarized photons (negative $q$) to mostly come from the top and bottom quadrants while vertically polarized photons (positive $q$) will come mostly from the right and left.

\begin{figure*}
\centering
\includegraphics[width=1 \textwidth]{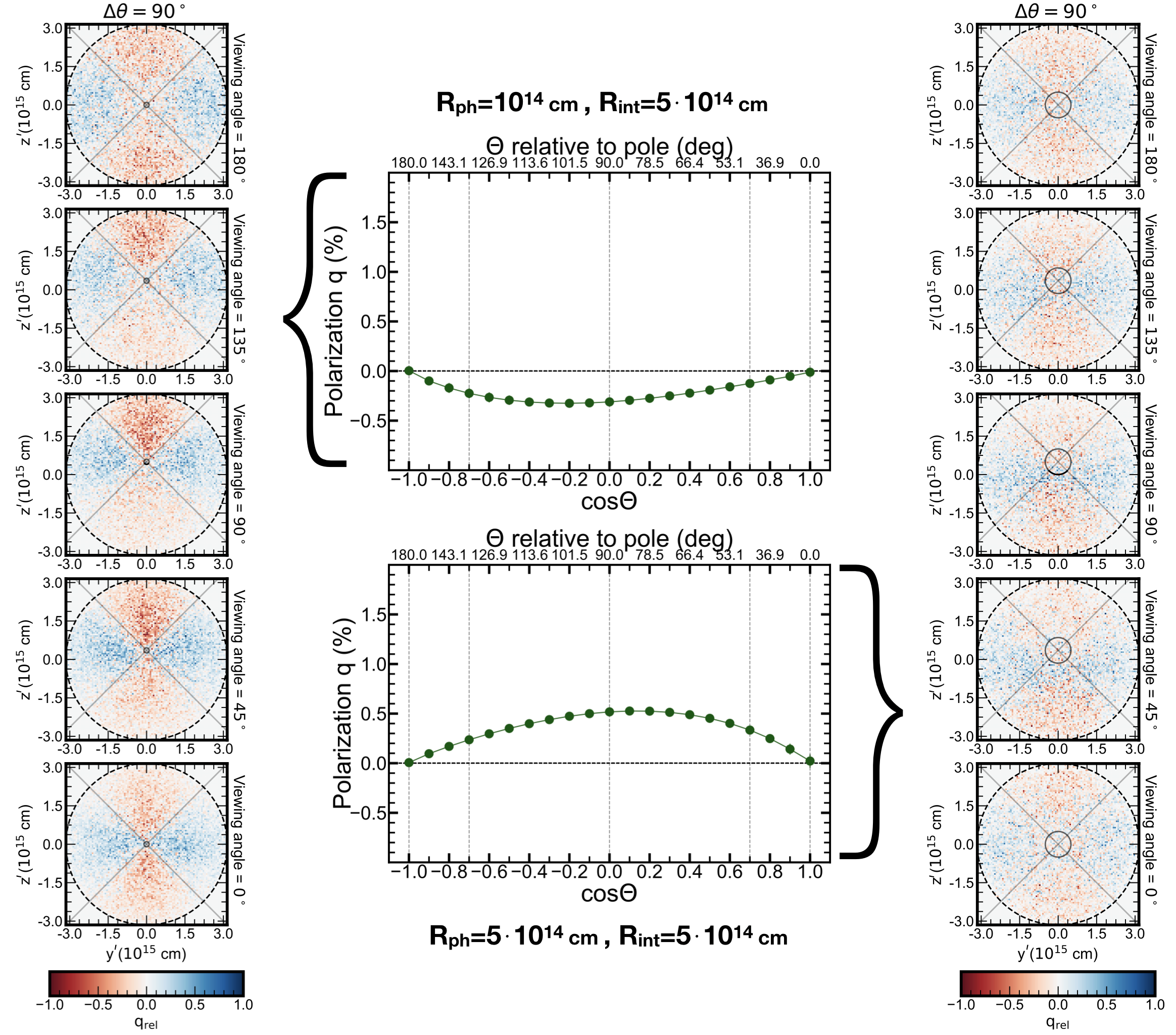}
\caption{Case A, $\mdotp = 2.93 \,\msun\, \rm yr^{-1}$. Investigating the polarization as a function of the viewing angle $\Theta$ for two simulations: s1; $\rphs$, $\rintl$, $\delthm$ (left column and top panel of middle column) and s2; $\rphl$, $\rintl$, $\delthm$ (right column and bottom panel of middle column). The middle column plots show the polarization as a function of the viewing angle for each particular simulation (identical with Fig. \ref{fig:med_mass_subs}). The vertical dashed lines denote five different viewing angles which correspond to each subplot of the left and right column. The left and right column subplots show the contribution of each single photon in the final polarization value (relative normalized polarization colormap values) for five different viewing angles ($\Theta$) which from top to bottom are: $\Theta=180^{\circ},135^{\circ},90^{\circ},45^{\circ}\, {\rm and}\, 0^{\circ}$. Red color denotes a negative $q$ (i.e. oscillation of the electric field in the horizontal direction) and blue color denotes a positive $q$(i.e. oscillation of the electric field in the vertical direction). The solid line circle in these subplots denotes the projection of the photosphere with radius $\rph$ while the black dashed line circle denotes the projection (for each observer) of the outflow (and the grid's) outer radius $\rmax$. The electron scattering photosphere $\dsc$ is not plotted in this case simply because the density in the outflow is so high that $\tau>1$ everywhere inside $\rmax$. Since each subplot is a 2D projection for a specific viewing angle, there is a rotation of coordinates in the $xz$ plane. Hence $y' = y$ and $z' = z\cdot\cos{\phi} - x\cdot\sin{\phi}$ where $\phi$ is the rotation angle which corresponds to $\phi=-90^{\circ},-45^{\circ},0^{\circ},45^{\circ}\, {\rm and}\, 90^{\circ}$ for each respective viewing angle $\Theta$. The four quadrants are plotted only as a visual aid.}
\label{fig:med_mass_explain}
\end{figure*}

\begin{figure*}
\centering
\includegraphics[width=1 \textwidth]{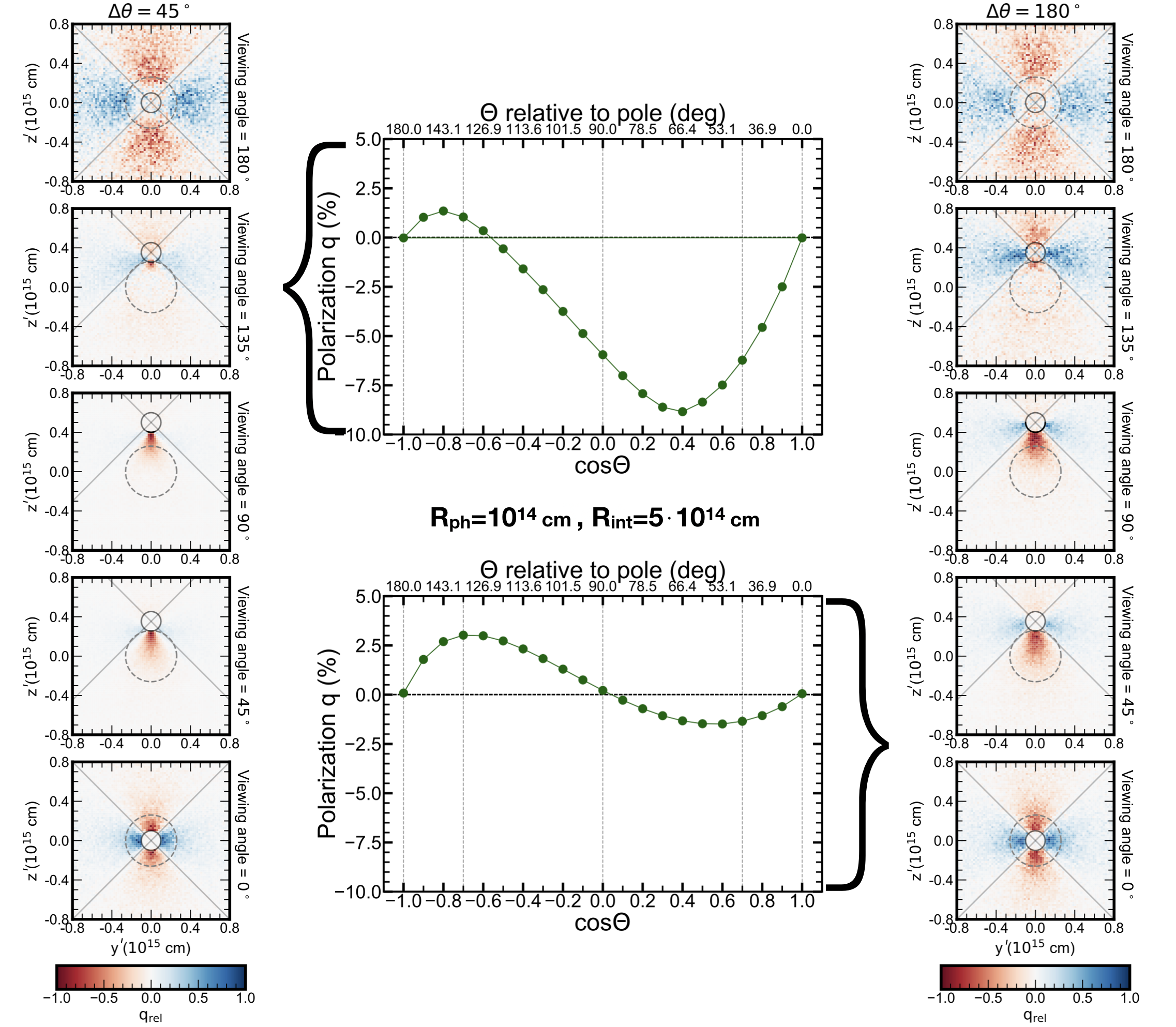}
\caption{Same as Fig. \ref{fig:med_mass_explain} but for Case B, $\mdotp = 0.29 \,\msun\, \rm yr^{-1}$, and for simulations: s3) $\rphs$, $\rintl$, $\delths$ (left column and top panel of middle column) and s4) $\rphs$, $\rintl$, $\delthl$ (right column and bottom panel of middle column). The grey dashed line circle here denotes the projection for each observer of the electron scattering photosphere with radius $\dsc$.}
\label{fig:low_mass_explain}
\end{figure*}

Based on the above, we attempt to explain the different results derived from our simulations and for that, we choose two cases that are quite different from each other, a pair of cases for Case A (Fig. \ref{fig:med_mass_explain}) and a pair for Case B (Fig. \ref{fig:low_mass_explain}). For each of these individual cases, we store the $x,y,z$ coordinates and $q$ value at the location of the last scattering of each individual photon before it leaves the system to reach the observer in the following viewing angles ($\Theta$): $\Theta=0^{\circ},45^{\circ},90^{\circ},135^{\circ}\, {\rm and}\, 180^{\circ}$. Then, for each individual case, we divide those $q$ values by the maximum $|q|$ of the five $\Theta$ (relative normalised polarization, $q_{rel}=q/|q_{max}|$) and we plot it as a colormap, on top of the rotated $y$ and $z$ coordinates ($y' = y$ and $z' = z\cdot\cos{\phi} - x\cdot\sin{\phi}$ where $\phi$ is the rotation angle which corresponds to $\phi=-90^{\circ},-45^{\circ},0^{\circ},45^{\circ}\, {\rm and}\, 90^{\circ}$) for each respective viewing angle $\Theta$. Since the number of photons stored in those files is very large, we average the $q$ contributions within a spatial bin range, varying for every case depending on the number of photons stored.

The comparison for Case A is presented in Fig. \ref{fig:med_mass_explain} where we focus on two simulations, which we call s1 and s2. Simulation s1 has $\rphs$ and $\rintl$ and is shown on the top and left subplots, while s2 has $\rphl$ and $\rintl$ and is shown on the bottom and right subplots. Both simulations have the same opening angle $\delthm$. 
For the first case (s1),
contributions come mostly from the top quadrant 
while much fewer photons are scattered with a clear direction from the remaining quadrants as these are closer to the intersection point and thus at higher densities. 
Photons from the top quadrant are preferentially polarized in the horizontal direction and manage to leak from the system (after several scatterings) in the less dense path. Most leave from the top quadrant and only a small fraction of the photons strongly scattered near the intersection point diffuse towards the left and right quadrants.The resulting polarization is dominated by the top quadrant and therefore by a negative $q$. For $\Theta=45^{\circ}$, positive and negative contributions almost balance each other and as we move towards $90^{\circ}<\Theta<135^{\circ}$ the negative ones slightly prevail, making the peak of the polarization be offset towards those larger viewing angles. In the second case (s2), $\rphl$ and $\rintl$, positive contributions prevail for two main reasons. First, in cases where $\rph\approx\rint$, photons are injected very close to the very dense area around the intersection point and immediately undergo many scatterings. After they scatter, if a photon enters the emitting photosphere, complete thermalization is assumed and the photon is removed from the simulation. Hence, in such cases, way less photons manage to escape the system. This is depicted in the faint-colored colormap as way less contributions are stored for this case. Now, since the emitting surface of the photosphere is five times closer to the center of the grid, photons are not emitted only deep into the top quadrant (like they do in s1). Combined with the fact that the emitting surface is five times larger, photons are not (initially) directed only towards the very dense center, but many of them are first injected into and then diffuse towards the left and right quadrants, to eventually emerge from the system with a positive $q$ value (electric field oscillating in the vertical direction).

A similar study for Case B is presented in Fig. \ref{fig:low_mass_explain} where we focus on two simulations of $\rphs$ and $\rintl$; one with $\delths$ (s3) and one with $\delthl$ (s4). For both simulations, a viewing angle of $\Theta=135^{\circ}$ (i.e. looking closer to the side of the intersection point than the BH) results into positive polarization (i.e. vertical oscillation). Much fewer photons reach the observer along this line of sight as the very dense intersection point starts getting between the emitted photons and the observer. That leads to very few contributions from the lower quadrant and contributions from the left and right (positive) end up being stronger than the ones from the top quadrant (negative). Hence the signal for $\Theta=135^{\circ}$ is positive. 

In general, for case B, the optical depth drops to order unity such that photons are scattered towards the observer shortly after being injected at the photosphere. For viewing angles close to $\Theta=180^{\circ}$, the region of the photosphere where photons are injected intersects with the left and right quadrants, leading to the positive polarization value seen in s3 and s4 of Fig. \ref{fig:low_mass_explain}. As the viewing angle $\delth$ decreases below a critical value, all the photons are instead injected in the bottom quadrant, causing the polarization degree to become negative. As expected geometrically, this critical viewing angle is lower when the opening angle increases, as can be seen by comparing the top and bottom panels. As $\delth$ becomes larger, the contribution of the left and right quadrants increases leading to a change in polarization value, from $q\approx1.5\%$ for $\delths$, to $q\approx3\%$ for $\delthl$. For $\Theta=45^{\circ}\;\rm{or}\;90^{\circ}$ we have a strong contribution from the lower quadrant. For $\delths$ the photons are emitted strictly to the bottom quadrant leading to negative polarization (i.e. horizontal oscillation) values for observers at these viewing angles. As we move to larger $\delth$ (i.e. a larger photospheric surface from which the photons are injected into the system), contributions start coming from the left and right quadrants equalizing the strong contribution of the bottom quadrant. This eventually changes the polarization value for viewing angles $45^{\circ}<\Theta<90^{\circ}$ (i.e. observers viewing the system from the hemisphere containing the BH ), from $P\approx0-1.5\%$ for s4 to the striking $P\approx6-8.8\%$ for s3.

An interesting comparison in order to try and understand better what affects the emerging polarization value for a specific viewing angle is to compare cases s1 to the s3 and s4, for viewing angles around $\Theta\sim135^{\circ}$ for which, in principle, we see the same effect. That is, the very dense area around the intersection point enters the line of sight and photons get highly scattered, trying to find the less dense path and eventually escape the system. These are the top, left and right quadrants and rarely the bottom quadrant. For s1 they seem to do that more successfully from the top quadrant (with horizontal scatterings) resulting to negative $q$ while for s3 and s4 they do that more successfully from the left and right quadrants combined (with vertical scatterings) resulting to positive $q$. This happens because in s1, photons are injected into a very dense environment ($\tau\gg1$) and photons manage to escape from the path that provides even the slightest drop in optical depths (which is in the top quadrant). For s3 and s4 the environment is not that optically thick ($\tau\lesssim1$, see in Fig. \ref{fig:low_mass_explain} that the electron scattering photosphere is much closer to the injection surface) hence they manage to leak from the system from the left and right quadrants as well (with horizontal scatterings and positive $q$).


Unfortunately, the parameter space (e.g. of optical depths/densities, potential geometrical setups and inclination angles) is vast and this makes interpretations of observational data subject to degeneracies.
The careful study, however, of data and figures such as Figs \ref{fig:med_mass_explain} and \ref{fig:low_mass_explain}, can provide us with important insights and better understanding of such intricate processes. In Fig \ref{fig:scat} of the Appendix, we present an extra test on what might dominate the polarization. Using the same file that was used to create the colormaps of s2 in Fig. \ref{fig:med_mass_explain} and of s3 in Fig. \ref{fig:low_mass_explain}, we reproduce the polarization curves but by every time removing the contribution of those photons that scattered once, then those scattered once or twice etc, until reaching a larger number of scatterings (10 or less scatterings). For the high density simulation of s2, we find that removing photons that scatter 10 or less times minimally affects the result implying that the polarization level is dominated by the multiple-scattered photons. For the case of simulation s3 of Fig. \ref{fig:low_mass_explain}, we find that in the lower density regime of Case B, the polarization is dominated by emerging photons that experienced $\lesssim5$ scatterings. Furthermore, we find that there is a viewing angle dependence and can be seen by the fact that the positive $q$ values around the viewing angles of $\Theta=135^{\circ}$ are set by photons that scattered only once.

\subsection{Time evolution} \label{subsec:results_time}

\begin{figure*}
\centering
\includegraphics[width=1 \textwidth]{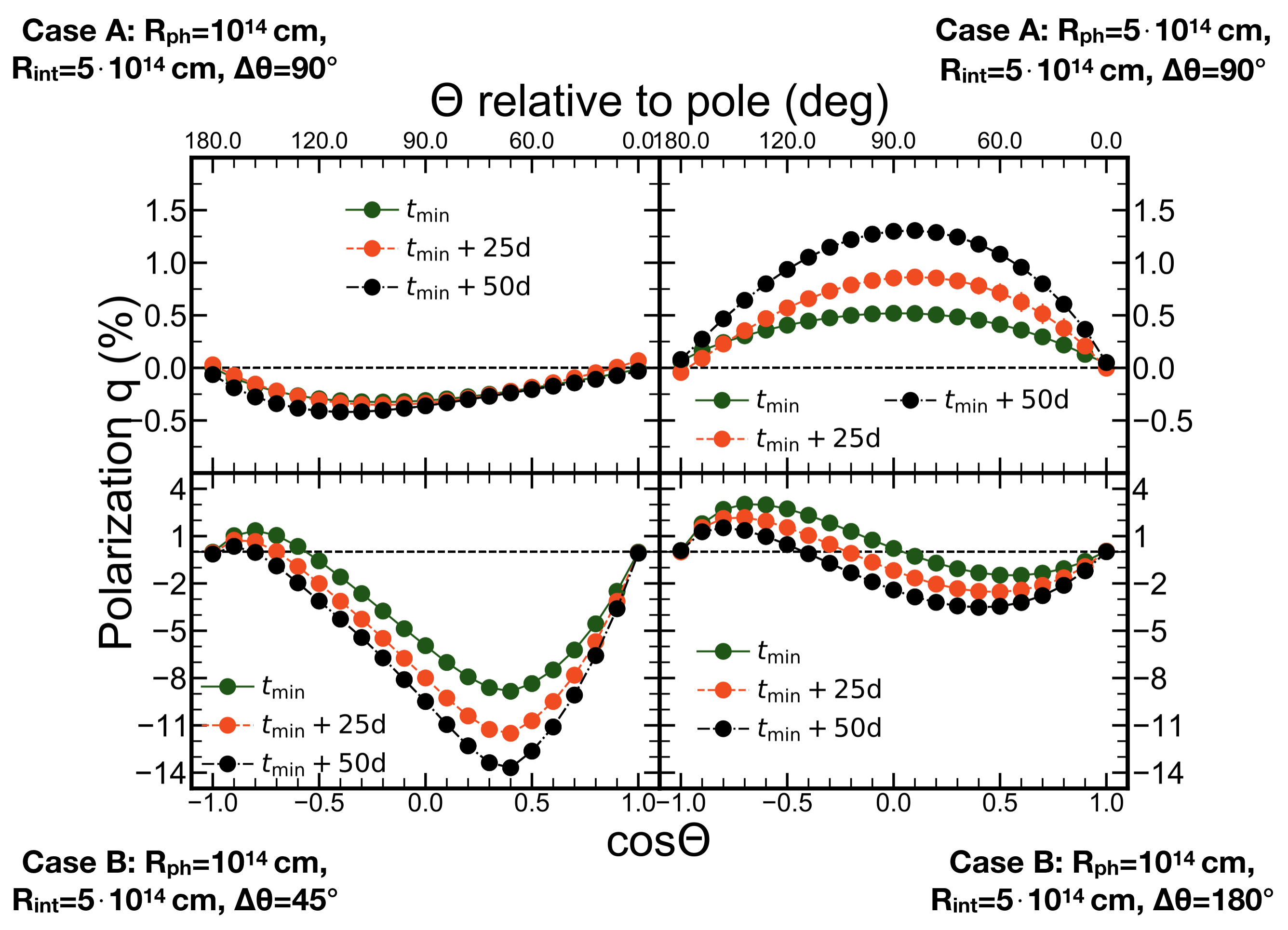}
\caption{Time evolution of the polarization as a function of the viewing angle. Each one of the four subplots of this plot probes a specific simulation setup, which is those discussed in Figures \ref{fig:med_mass_explain} and \ref{fig:low_mass_explain}. s1 and s2 (of Figure \ref{fig:med_mass_explain}) are in the top left and right panels respectively while s3 and s4 (of Figure \ref{fig:low_mass_explain}) are in the bottom left and right panels respectively. The different colors represent a time snapshot of this specific simulation, with green being the (already presented) t=$\tmin$d, orange being the t=$\tmin$+25d and black being the t=$\tmin$+50d. In general, we do observe a change of polarization with time and the time evolution from $\tmin$ seems to favour higher maximum $P$ values. However, it could result in lower observed $P$ values for observers at specific viewing angles.}
\label{fig:time_snap}
\end{figure*}

The simulations presented in Sect. \ref{subsec:results} provide polarization predictions for two different mass outflow rate cases. In order to study how the polarization signal from a TDE evolves with time, we ran simulations for two more epochs, at 25 and 50 days after $\tmin$. Hence, the density of such grids is provided by Equation \eqref{eq:density_time} at snapshots of $t=t_{\rm min}$+25d and $t=t_{\rm min}$+50d respectively. We ran those snapshots only for the extensively studied simulations presented in Figures \ref{fig:med_mass_explain} and \ref{fig:low_mass_explain}, 
i.e. s1, s2 (from the high mass outflow rate Case A) and s3 and s4 (from the low mass outflow rate Case B). We present the results in Fig. \ref{fig:time_snap}.
The main effect of time evolution is that the densities and optical depths in the grid decrease (see discussion on Sect. \ref{subsubsec:optdsandvas} and Table \ref{tab:densandtau} for indicative optical depths and densities at various distances $r$ from the intersection point, for the three different time snapshots). This is a natural outcome of the expanding spherical volume of the outflow since its radius expands with a constant velocity $\vout$ even though the embedded mass $M_{\rm env}$ is also gradually increasing (see Table \ref{tab:sims} for detailed numbers). The time evolution seems to generally favour higher maximum $P$ values for all the studied cases. However, time evolution could also result in lower observed $P$ values for some observers at specific viewing angles. For example, if we focus on the low-mass case of s4, an observer around $\Theta\sim120^{\circ}$ would observe the polarization signal to steadily drop with time (from $P\simeq2.5\%$ to $P<0.5\%$), but an observer around $\Theta\sim60^{\circ}$ would see the signal to rise (from $P\simeq1.5\%$ to $P<3.5\%$). This opposite trend for different observers, is present in the low-mass cases even for some observers at neighbouring viewing angles; taking the example of s3 now, an observer at $\Theta\sim127^{\circ}$ ($\cos\Theta=-0.6$) would see the the polarization signal to rise from $P\simeq0\%$ to $P\simeq2\%$ while an observer at $\Theta\sim143^{\circ}$ ($\cos\Theta=-0.8$) would see the the polarization signal to drop from $P\simeq1.4\%$ to $P\simeq0\%$. These degeneracies make associating an observation of a TDE with a specific viewing angle hard, even if multiple epochs are available. 

The fact that the signal increases for s2 is natural, since the optical depths drop and photons do not undergo multiple scatterings which in turn cause a loss of information on directionality and destruction of the polarization signal. However, for s1, we practically see no change in the polarization levels in a span of 50 days. In this simulation photons are emitted further away from very dense intersection point compared to s2. Hence for s1 the change in density and optical depths as time passes, do not make a difference yet. However for s2, where photons are injected in very small distances $r$ from the center making the densities steeply increase (since $\rho_{\rm out}\propto r^{-2}$; see Equation \eqref{eq:density_time}), the drop in densities that come with the pass of time does make a difference and we see polarization rising for all the viewing angles. We note that, from every setup, as time evolves and the environment becomes very optically thin ($\tau\ll1$),
it is expected that photons will eventually start escaping without scattering on electrons. That will lead to a drop in polarization values.

The effects that varying mass, density and optical depths has in the polarization signal are studied and attempted to interpret thoroughly in Sect. \ref{subsec:results} as well as in the discussion of this work (Sect. \ref{sec:discussion}).


\section{Discussion} \label{sec:discussion}
\subsection{What properties affect the polarization?} \label{subsec:disc1}

Our parametric study showed that the observed polarization level primarily depends on i) the optical depth (or density) in the central regions where the photons are emitted (which
depends on the mass outflow rate $\mdotout$), and ii) the viewing angle. However, due to the large parameter space, multiple degeneracies in the $P$ value arise and observed values of polarization can be explained by several combinations of parameters.

\subsubsection{Dependence on the physical parameters} \label{subsubsec:optdsandvas}

\begin{figure}
\centering
\includegraphics[width=0.49 \textwidth]{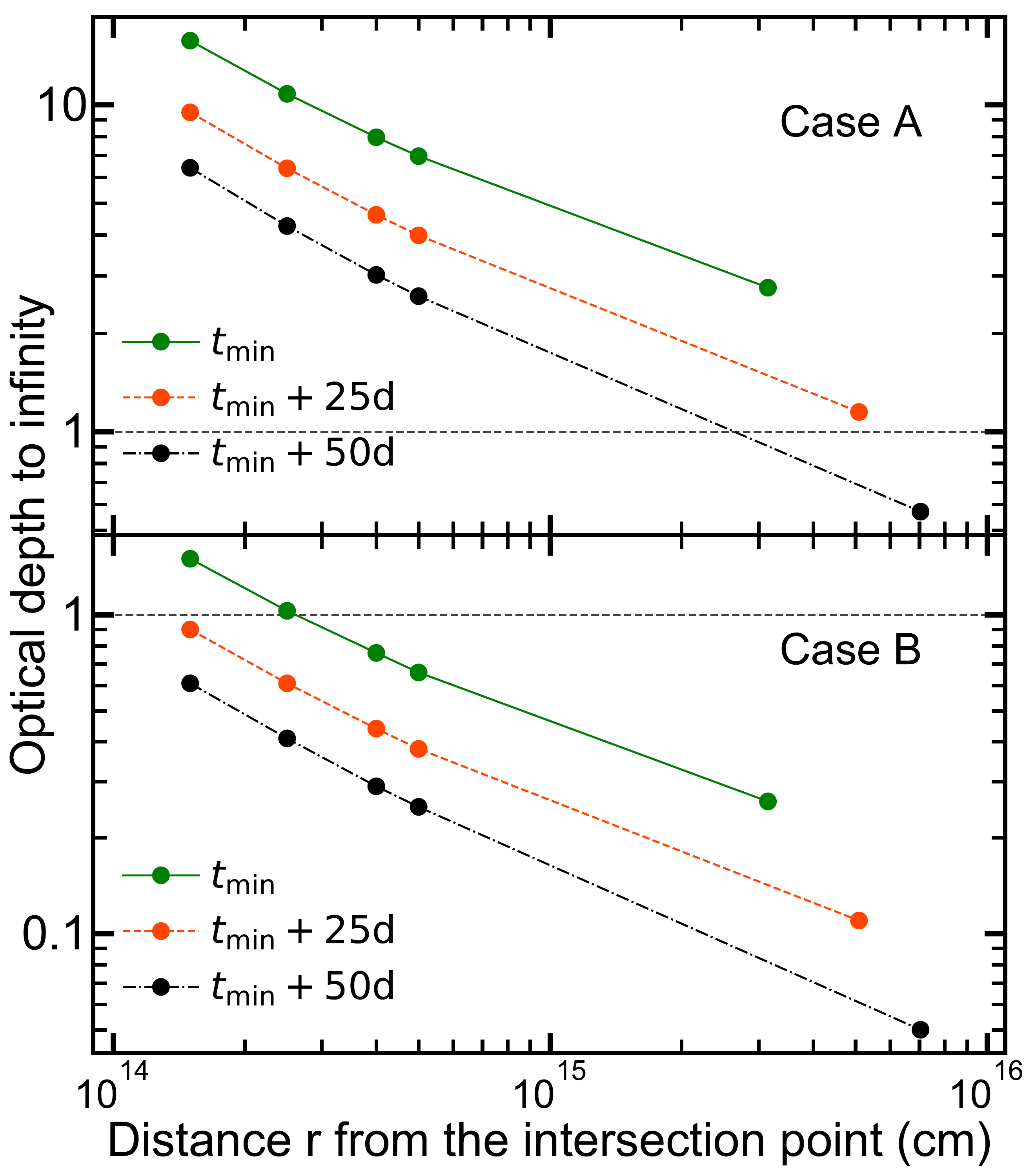}
\vspace{0cm}
\caption{Indicative optical depths from $r$, the distance from the intersection point, to infinity, for the two different mass outflow rate cases (Case A top panel and Case B bottom panel) and for three different time snapshots ($t=\tmin$, $t=\tmin+25\rm d$ and $t=\tmin+50\rm d$). The outermost right point of every curve is where the $\rmax$ is located, that is, the outer radius of the outflow/grid. The horizontal dashed lines is where the electron scattering photosphere $\dsc$ lies (by definition at $\tau=1$). The exact numbers shown in this plot can be found in Table \ref{tab:densandtau} in the Appendix.}
\label{fig:optds_vs_r}
\end{figure}

In Fig. \ref{fig:optds_vs_r}, we present indicative optical depths at various distances $r$ from the intersection point (center of the grid) for both Cases A and B and for three different time snapshots ($t=\tmin$, $t=\tmin+25\rm d$ and $t=\tmin+50\rm d$). A similar figure with the densities as a function of $r$, as well as a table with the exact numbers presented in these two figures, can be found in Fig. \ref{fig:dens_vs_r} and Table \ref{tab:densandtau} in the Appendix. As the mass becomes higher, the central region around the intersection point (close to the photosphere surface from where the seed photons are injected), becomes very dense (optically thick). This very high density leads to multiple scatterings in every direction which in turn leads to a decrease of polarization. As the density drops (as in Case B or with time) the optical depth around the intersection point drops as well and polarization values starts rising. 
For Case A, we see that the only simulations where the polarization is higher than $P>0.5\%$ is for the combinations of $\rphl$, $\rintl$ with $\delthm$ and $\delthl$. These are the cases (for Case A) for which photons are injected in the system as far as possible from the center (intersection point). The closest point where photons are injected for these cases is of course at $\rint-\rph=0$. With simple geometrical calculations we can find that for the first case, the furthest distance at which photons are emitted is at $r=7.07\times10^{14}\rm cm$, and for the second case at $r=10^{15}\rm cm$. In Table \ref{tab:densandtau}, for Case A, we see that $r=5\times10^{14}\rm cm$ has an optical depth of $\tau=6.97$ and $r=\rmax=3.15\times10^{15}\rm cm$ an optical depth of $\tau=2.76$. Hence we find that when photons are injected in regions of optical depth below $\tau\ll7$ we start seeing polarization levels of $P\gtrsim0.5\%$. Else the environment is very optically thick and information on directionality is lost. Of course that condition alone is not enough to explain under which conditions we get some polarization but this is a simple quantification of one of the conditions needed to get some polarization signal.

\subsubsection{Dependence on the geometrical parameters} \label{subsubsec:freeparams}

Geometrical parameters such as the viewing angle and the free parameters of our model ($\rint$, $\rph$, $\delth$) eventually affect the polarization by causing changes in the optical depth.

The viewing angle affects the amount of material that is between the emitting source and the observer as well as the asymmetry in the projected geometry of the emitting source (e.g. the projected geometry of our model is always symmetrical for $\Theta=0^{\circ}$ and $\Theta=180^{\circ}$ but not for the other viewing angles). The predicted polarization signal shows a clear viewing angle dependence and for the vast majority of the simulations of the high mass outflow rate Case A, it seems that the absolute value of polarization reaches its maximum for equatorial viewing angles ($-0.4\leq\cos\Theta\leq0.4$ or $114^{\circ}\leq\Theta\leq66^{\circ}$). On the other hand, for the low mass outflow rate Case B, we see the highest values for closer to polar viewing angles ($ 0.4\leq|\cos\Theta|\leq 0.7$ or $45^{\circ}\leq\Theta\leq66^{\circ}\,\&\,114^{\circ}\leq\Theta\leq135^{\circ}$). Since our model is axially symmetric, the polarization for both polar views is zero as expected. Hence we show that an alignment of i) observer, ii) black hole and iii) intersection point or i) observer, ii) intersection point and iii) black hole, would result in no observable polarization signal (as expected since for polar viewing angles our system is symmetric).

The free parameters of our model dictate where the photons are injected in the system and different locations mean different densities and optical depths. The dependence on the free parameters of our model ($\rint$, $\rph$, $\delth$) is discussed in Sect. \ref{subsec:results} and demonstrated for two pairs of simulations in Figs. \ref{fig:med_mass_explain} and \ref{fig:low_mass_explain}. In order to study this dependence in more depth and study how the polarization is affected by the variation of each individual parameter, we present the following plots whose upper panels refer to Case A and lower panels refer to Case B.

\begin{figure}
\centering
\includegraphics[width=0.49 \textwidth]{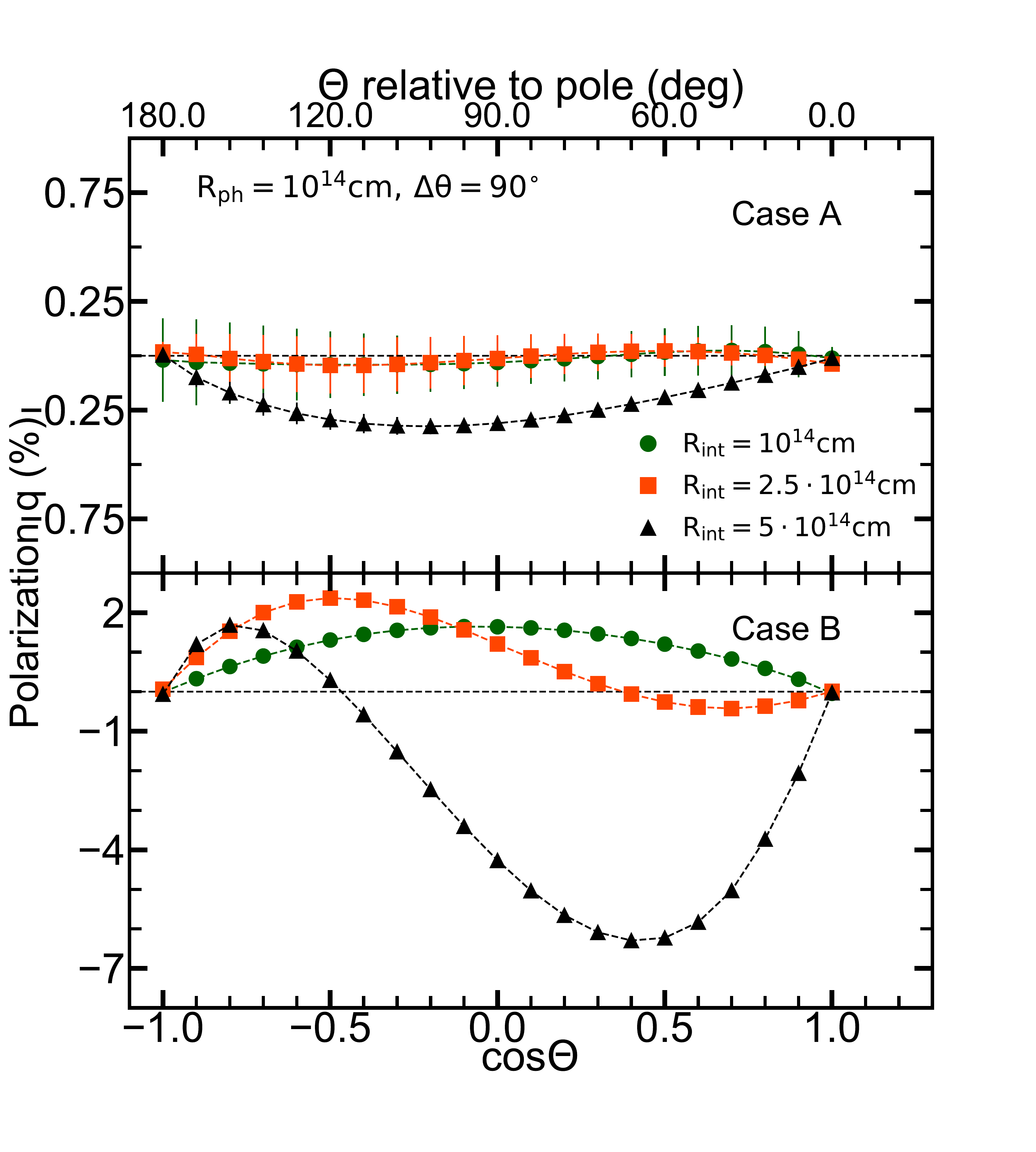}
\vspace{-1cm}
\caption{Investigating the polarization as a function of the viewing angle $\Theta$ for varying $\rint$ while we keep $\rph$ and $\delth$ fixed. The upper panel is for the high mass Case A and the lower panel for the low mass Case B. Extending the distance $\rint$ between the intersection point and the black hole seems to generally favour higher polarization levels for most viewing angles.}
\label{fig:rints}
\end{figure}

In Fig \ref{fig:rints}, we study how the variation of $\rint$ affects the polarization. Therefore we keep the radius of the photosphere and the opening angle fixed to $\rphs$ and $\delthm$ and we vary the intersection radius ($\rints$, $\rintm$ and $\rintl$). We find that as $\rint$ becomes larger the maximum absolute value of the polarization ($P_{\rm max}$) becomes larger as well. For the high mass Case A (upper panel) we start getting some signal ($0\%<P<0.5\%$) only for $\rintl$ and that is because photons are now injected in a less optically thick environment than the other two cases (see discussion on s1, the upper panel of Fig. \ref{fig:med_mass_explain}). This change in $\rint$ and in turn in the optical depths around the injection surface, allows the photons to escape without all information on directionality being lost (as it happens for the two lower $\rint$ cases). For the low mass Case B that has generally lower optical depths (lower panel), apart from the larger $P_{\rm max}$, the change of $\rint$ produces also a strong viewing angle dependence.
We see the viewing angle of $P_{\rm max}$ to shift from being close to the equator ($\Theta\approx90^{\circ}$), to
angles $\approx60^{\circ}$ from the poles.
A further complication is that, depending on $\rint$, the maximum polarization can be observed for viewing angles either above or below the equator (from the same or the opposite hemisphere including the BH) and it can also change signs (i.e. the observer switches from measuring a vertical to a horizontally oscillating electric field).

\begin{figure}
\centering
\includegraphics[width=0.49 \textwidth]{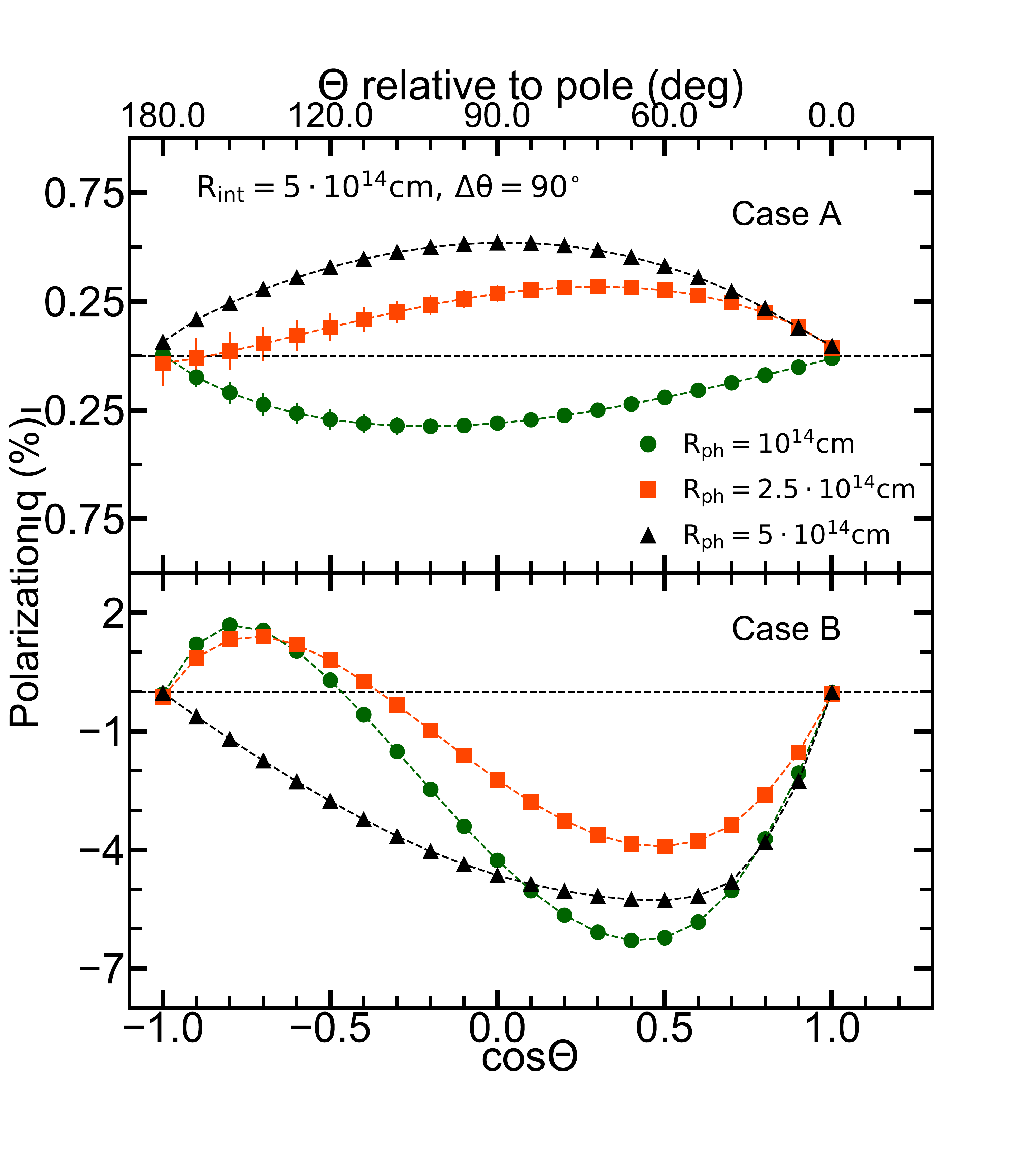}
\vspace{-1cm}
\caption{Investigating the polarization as a function of the viewing angle $\Theta$ for varying $\rph$ while we keep $\rint$ and $\delth$ fixed. The upper panel is for the high mass Case A and the lower panel for the low mass Case B. The dependence of polarization with varying $\rph$ is less trivial.}
\label{fig:rphs}
\end{figure}

In Fig. \ref{fig:rphs}, we study how the variation of $\rph$ affects the polarization. Therefore we keep the intersection radius and the opening angle fixed to $\rintl$ and $\delthm$ and we vary the photospheric radius ($\rphs$, $\rphm$ and $\rphl$). For the high mass Case A (upper panel) we find that as $\rph$ becomes larger, the polarization signal shifts from being negative to becoming positive. This happens because, as the photospheric radius becomes larger, photons are injected closer to the very dense center and preferentially escape from the less dense path of the left and right quadrants and are thus polarised in the vertical direction (i.e., positive $q$). For the low mass Case B (lower panel) the evolution of the signal with varying $\rph$ is less trivial. We see that, depending on the viewing angle, the observer switches from measuring a vertical to a horizontally oscillating electric field (this effect is thoroughly discussed in Fig. \ref{fig:low_mass_explain} for s3 and s4). While the $q$ values for viewing angles close to face-on become lower going from $\rphs$ to $\rphm$, they rise again for $\rphl$, where all the signal becomes negative. This change is probably an effect of $\rph$ becoming equal to $\rint$ (see discussion on s2, the bottom panel of Fig. \ref{fig:med_mass_explain}).

The variation of the opening angle $\delth$ seems to have different effect on the polarization depending on the distances between $\rph$ and $\rint$. For this reason, we probe two different combinations of $\rint$ and $\rph$ while we vary the opening angle ($\delths$, $\delthm$, $\delthl$) and we show the results in the left and right panels of Figure \ref{fig:delths} in the Appendix.

To summarize, increasing the distance $\rint$ between the intersection point and the black hole seems to generally favour higher polarization levels. This happens because as we move away from the very dense/optically thick intersection point, it is easier for photons to escape with fewer scatterings, so there is no loss of directionality. Also, widening the opening angle $\delth$ when $\rph$ and $\rint$ are similar (like the example in the right panel of Fig. \ref{fig:delths}) seems to favour higher polarization levels as well. This happens due to a combination of factors. The bottom quadrant (see e.g Figure 4) is hard for photons to leak to and escape from, because it is where the very dense/optically thick intersection point is. This gives an advantage to positive polarization as contributions from left and right quadrants overcome those of the top one. However if the distance between $\rph$ and $\rint$ becomes larger, widening the opening angle $\delth$ seems to favour lower polarization levels. This happens because photons are not radiated in a very optically thick environment and we get negative polarization from the bottom quadrant (like in the examples of Fig. \ref{fig:low_mass_explain} and the left panel of Fig. \ref{fig:delths}). However in higher optical depth cases this changes, (e.g. top panel of Fig. \ref{fig:med_mass_explain}; Case A), because material is optically thick in the area that photons are injected so there is no signal from the bottom quadrant even though the intersection radius is large.

\subsection{Connection with observations} \label{subsec:disc2}

A key goal of polarization modeling is to enable us to constrain the viewing angle (and other physical properties) of actual observations. In this Section, we compare our predicted continuum polarization values with polarization observations of three different TDEs presented in \citet{Leloudas2022}; AT~2018dyb, AT~2109dsg and AT~2019azh, and the one presented in \citet{Patra2022}; AT~2019qiz.

\begin{figure}
\centering
\includegraphics[width=0.49 \textwidth]{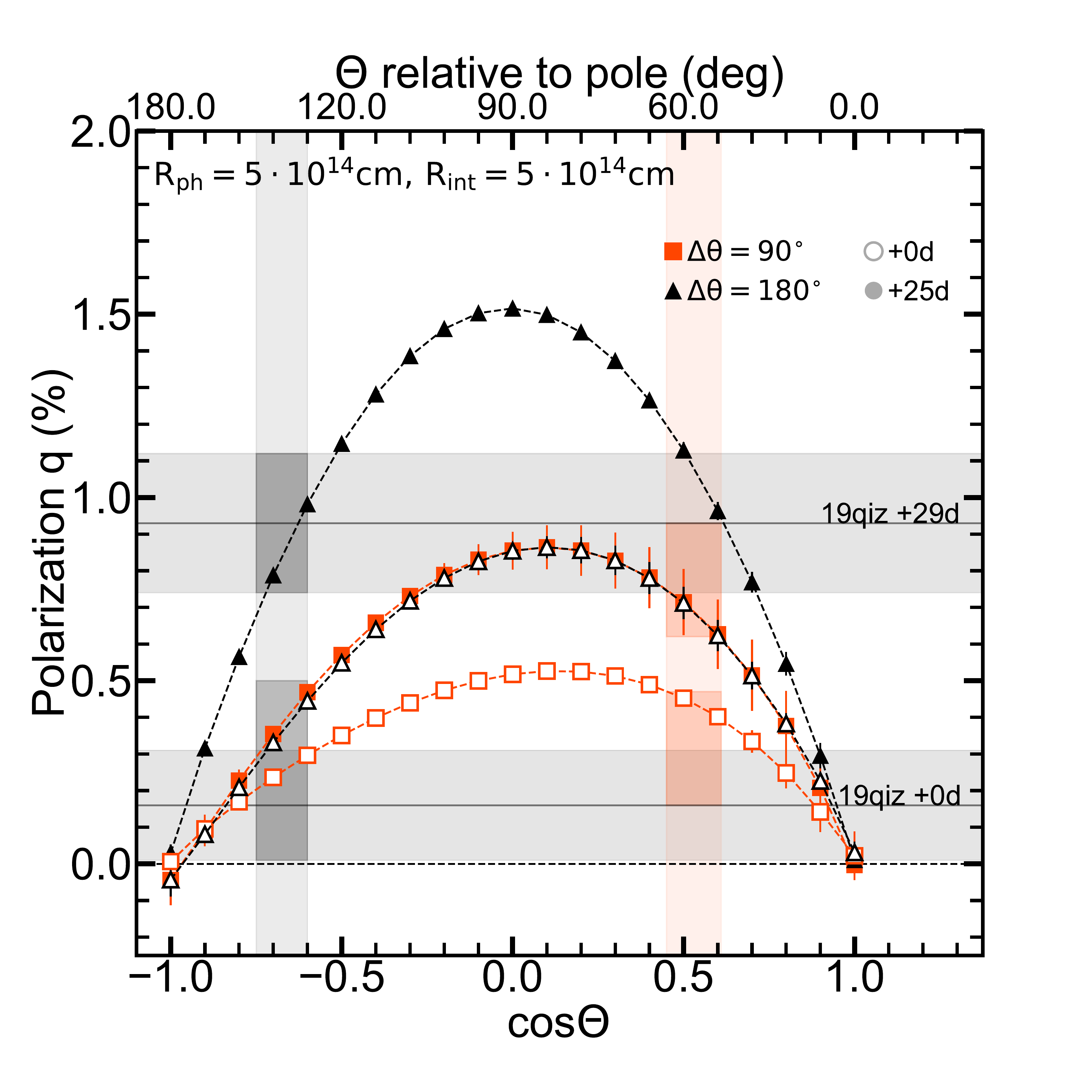}
\vspace{-1cm}
\caption{Polarization $q$ as a function of the viewing angle for two Case A simulations of $\delthm$ and $\delthl$ with $\rphl$ and $\rintl$. The $\mdotp$ of Case A ($\mdotp=2.93\, \msun\, \rm yr^{-1}$) is similar with the one that we calculate for AT~2019qiz ($\mdotp=2.79\, \msun\, \rm yr^{-1}$). Since there are polarimetric observations of AT~2019qiz at +0d and +29d with respect to peak \citep{Patra2022}, we present our aforementioned simulations at $t=\tmin$ and $t=\tmin+25\rm d$ in order to make a realistic comparison. The horizontal solid grey lines are the $P$ values observed and the shaded area around it is the uncertainty of the measurements. The shaded vertical areas are potential viewing angles that can predict successfully the evolution of the polarization of the TDE. The more opaque regions of these shaded vertical areas are just a visual aid to guide the eye; they show where the simulated $P$ values for the different time snapshots almost coincide with the observed $P$ values, and their color matches the color of each simulation setup.}
\label{fig:19qiz}
\end{figure}

\begin{figure}
\centering
\includegraphics[width=0.49 \textwidth]{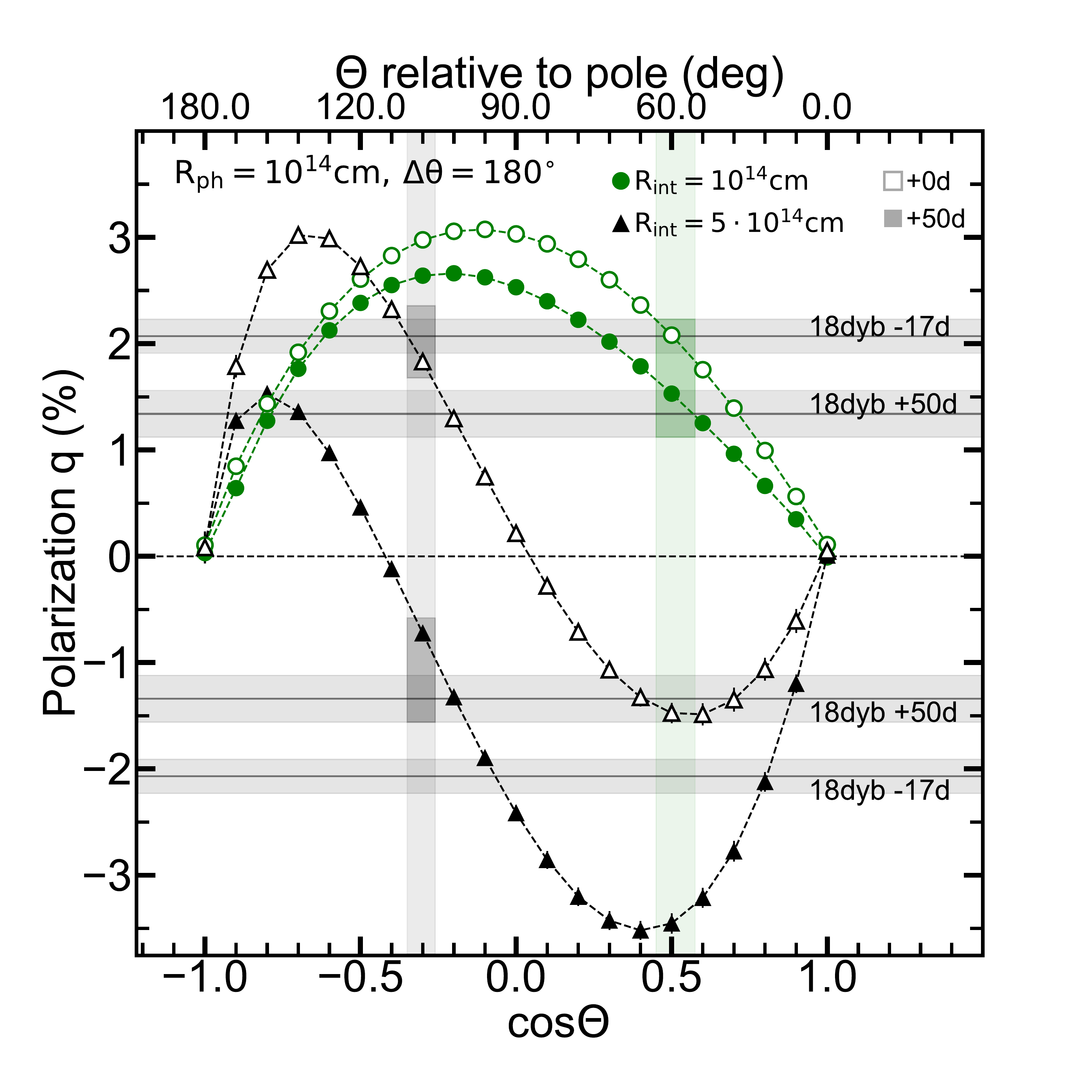}
\vspace{-1cm}
\caption{Polarization $q$ as a function of the viewing angle for two Case B simulations of $\rints$ and $\rintl$ with $\rphs$ and $\delthl$. The $\mdotp$ of Case B ($\mdotp=0.29\, \msun\, \rm yr^{-1}$) is similar with the one that we calculate for AT~2018dyb ($\mdotp=0.16\, \msun\, \rm yr^{-1}$). Since there are polarimetric observations of AT~2018dyb at -17d and +50d with respect to peak \citep{Leloudas2022}, we present our aforementioned simulations at $t=\tmin$ and $t=\tmin+50\rm d$ in order to make a realistic comparison. The horizontal solid grey lines are the $P$ values observed and the shaded area around it is the uncertainty of the measurements. The shaded vertical areas are potential viewing angles that can predict successfully the evolution of the polarization of the TDE. The more opaque regions of these shaded vertical areas are just a visual aid to guide the eye; they show where the simulated $P$ values for the different time snapshots almost coincide with the observed $P$ values, and their color matches the color of each simulation setup.}
\label{fig:18dyb}
\end{figure}

Degeneracies between the mass outflow rates and the viewing angles make it hard to constrain a single viewing angle for an observed TDE (for a simulation at a time snapshot that coincides with the phase of the observation). In order to break these degeneracies, we have estimated the (peak) mass outflow rate ($\mdotp$) of each TDE, using Equation \eqref{eq:calcs}. To obtain our best estimates we have used the $\mh$ and the $\mstar$ masses derived by fitting the TDE light curves with MOSFiT \citep{Nicholl2022}. We determined the $\rstar$ as function of their mass exactly like \citet{Mockler2019} that is, we used the mass–radius relation for ZAMS solar-metallicity stars given in \citet{Tout1996} for main-sequence stars. 

\begin{table}
\renewcommand{\arraystretch}{1.2}
\setlength\tabcolsep{0.1cm}
\fontsize{8.8}{11}\selectfont
\centering
 \caption{Indicative $\mdotp$ values (last column) of the four TDEs that we compare our results to. The $\mh$ (second column) and $\mstar$ (third column) have been retrieved from \citet{Nicholl2022}. $\rstar$ (fourth column) is calculated as function of the $\mstar$ using the mass–radius relation for ZAMS solar-metallicity stars given in \citet{Tout1996} for main-sequence stars. }
 \label{tab:mdotp}
 \begin{tabular}{c|cccc}
  \hline
  TDE & $\log({\mh/\msun})$ & $\mstar/\msun$ & $\rstar/\rsun$ & $\bf{\dot{M}_{\rm \bf{p}}/(\mathrm{\bf{M}}_{\hbox{$\bf{\odot}$}}\, \rm yr^{-1})}$\\
  \hline
  AT~2018dyb & 7.19 ($\pm$0.02) & 0.1 ($\pm$0.00) & 0.13 ($\pm$0.01) & \bf{0.16 ($\pm$0.02)} \\
  AT~2019dsg & 6.57 ($\pm0.19$) & 0.91 ($\pm$0.43) & 0.81 ($\pm$0.33) & \bf{1.78 ($\pm$2.13)} \\
  AT~2019azh & 6.7 ($\pm0.07$) & 0.47 ($\pm$0.14) & 0.43 ($\pm$0.12) & \bf{1.05 ($\pm$0.76)} \\
  AT~2019qiz & 6.22 ($\pm0.04$) & 1.01 ($\pm$0.03) & 0.90 ($\pm$0.03) & \bf{2.79 ($\pm$0.22)} \\
  \hline
 \end{tabular}\\
\end{table}

In Table \ref{tab:mdotp} we present our best estimates of the $\mdotp$ values of the four TDEs studied in \citet{Leloudas2022} and \citet{Patra2022}. These values should be seen as indicative and with large uncertainties since $\mdotp$ depends on the masses of the SMBH, $\mh$, and of the disrupted star, $\mstar$. It is well known that these quantities are very hard to constrain. We also performed the error propagation of the uncertainties of these quantities and AT~2019dsg has a $\sigma(\mdotp)/\mdotp\sim200\%$ making this estimation unreliable while AT~2019azh has a $\sigma(\mdotp)/\mdotp\sim72\%$ making it very uncertain. For this reason we focus on AT~2018dyb and AT~2019qiz which have more trustworthy estimations of $\mdotp$. 
As we showed in this work, simulated polarization levels depend sensitively on $\mdotp$ so it is critical to better constrain $\mdotp$ of observed TDEs. Additional observables could be used in order to narrow down the number of potential viewing angles such as the emission or not of X-rays (and their strength) as well as distinct spectroscopic features, both of which have been used to predict the potential viewing angle that a TDE is seen from \citep{Dai2018,Charalampopoulos2022}. The following comparisons of simulations and observations should be taken as broad estimates and not as robust predictions since changes in the model parameters could affect the $q$ values of each viewing angle. However, taking into account the time evolution we can exclude many simulations (that is, values of the parameter space) as they are not able to reproduce the evolution of the signal. If we can reject many simulations, then we can potentially break the viewing angle degeneracies and make some broad estimates on potential viewing angles.

In order to compare observations and simulations, we have to make an assumption on how the phase of the observation of a TDE compares to our simulations' time snapshots. Since it still remains unknown when and what makes a TDE reach its optical luminosity peak, we consider that the peak of the optical light curve (defined as phase=0d for observations) is reached around $\tmin$ days after the fallback rate peak (t=0 in Equation \ref{eq:mdotfb}). This is a fair choice as the rise time (to the peak of the optical light curve) of $\tmin\approx40\rm d$ that is considered in this work (see Sect. \ref{sec:model}) is in accordance with observations and theoretical predictions (e.g. see \citealt{VanVelzen2020,Law-Smith2020}).

The estimated peak mass fallback rate of AT~2019qiz (${\mdotp=2.79\pm0.22}$) is similar to our Case A so we can compare the measurement of t=0d ($P=0.16\%$) with the results presented in our Fig. \ref{fig:med_mass_subs}. Case A was the case producing the lowest polarization values and the fact that AT~2019qiz has $P\sim0$ at peak \citep{Patra2022} is in agreement with the predictions of our model. In order to compare with measurements of $t=+29\rm d$ ($P=0.93\%$), we selected the simulations of Case A for the time snapshots of $t=\tmin+25\rm d$. We then sorted out which simulation setups can broadly reproduce \textit{both} the observed values, at the two different epochs. Even though the observation at $t=0\rm d$ ($P=0.16\%$) can be reproduced by virtually all the setups of $t=\tmin$, the $P=0.93\%$ at $t=\tmin+25\rm d$ can only be reproduced by two; two setups of $\delthm$ and $\delthl$ with $\rphl$ and $\rintl$. We visualise all the above in Fig. \ref{fig:19qiz}. The better match is clearly the simulation with $\delthl$ but we include the one with $\delthm$ as it could also predict the evolution of AT~2019qiz with small changes in the parameters. For the former, we find that the observables could be reproduced for observers at viewing angles of $127^{\circ}\lesssim\Theta\lesssim140^{\circ}$ (grey vertical shaded region in the plot). For the latter, we find that the observables could potentially be reproduced for observers at viewing angles of $53^{\circ}\lesssim\Theta\lesssim62^{\circ}$ (orange vertical shaded region in the plot). If we consider the better match case of $\delthl$, it suggests that the observations could happen from intermediate face-off viewing angles. That is a probable scenario since AT~2019qiz had mild X-ray emission ($L_{X}/L_{opt}\sim10^{-2}$; \citealt{Nicholl2020}) and X-rays could escape the system since $\delthl$ and not very narrowly directed to the intersection point, something that would reprocess almost all of them. The fact that $P$ rises to $\sim$1\% after one month from the peak could potentially be explained by the drop in density and optical depths, however it could be also explained by a change in geometry as \citet{Patra2022} suggest, or by a combination of both.
We caution that the observations of \citet{Patra2022} are not corrected for the effect of the host dilution from the host galaxy, which was shown to be very important by \citet{Leloudas2022}. While this would not affect the first observation of $P\sim 0\%$, the polarization at the second epoch can in fact be higher than $P=0.93\%$. Nevertheless, the idea behind our methodology still applies.

Case B has been the case associated with the highest predicted $P$ values and the fact that AT~2018dyb has the highest $P$ values of the TDEs observed to day, also is in agreement with the predictions of our model. AT~2018dyb has the lowest estimated peak mass fallback rate ($\mdotp=0.16\pm0.02$), a bit lower than our Case B so we can tentatively compare the measurement of $t=-17\rm d$ ($P=2.07\%$) with the results presented in our Fig. \ref{fig:low_mass_subs}. Admittedly, the polarization values of the snapshot at $t=\tmin$ might differ from the value of the $-17\rm d$ observation of AT~2018dyb however we include it as it can provide us with the trend of the polarization, that is, going from higher to lower values. In this way we can exclude simulations where the signal becomes larger with time. In order to fairly compare with measurements of $t=+50\rm d$ ($P=1.34\%$), we ran all the simulations of Case B for the time snapshots of $t=\tmin+50\rm d$. 
Then we sorted out which simulation setups can broadly reproduce both the observed values, at the two different time snapshots and we find that there are two setups of $\rints$ and $\rintl$ with $\rphs$ and $\delthl$. We visualise all the above in Fig. \ref{fig:18dyb}. Both simulations provide a good match but the simulation with with $\rints$ provides the better match and for this setup we find that the observed polarization could be reproduced for observers at viewing angles of $53^{\circ}\lesssim\Theta\lesssim63^{\circ}$ (green vertical shaded region in the plot). For $\rintl$, we find that the observables could potentially be reproduced for observers at viewing angles of $110^{\circ}\lesssim\Theta\lesssim115^{\circ}$ (grey vertical shaded region in the plot). All the above suggest that AT~2018dyb could have been observed from intermediate face-off and face-on viewing angles. We note though that it has been suggested in the literature that AT~2018dyb was observed from more equatorial viewing angles (e.g. \citealt{Leloudas2019,Charalampopoulos2022}) assuming however a different TDE emission mechanism scenario.

Finally, \citet{Law-Smith2020} provided a realistic library of fallback rates in TDEs and they calculate the expected mass fallback rates as a function of time for a large variety of combinations of stellar parameters (see their Figure 7). They find that the majority of full disruptions ($\beta$>1) of stars with $\mstar\geq1\,\msun$, should have an $\mdotp$ which is larger than three ($\mdotp\geq3\,\msun\, \rm yr^{-1}$). If our proposed scenario in this work is correct, then we expect that tidal disruptions of stars of $\mstar\geq1\,\msun$, should have continuum polarization levels below $P=1\%$ and for most of the cases below $P=0.5\%$.

\subsection{Limitations} \label{subsec:limits}

There are three main caveats in our approach that could make our model deviate from an absolutely realistic case.

1) It is very probable that the shape of the outflow is not perfectly spherical (shape of $\rmax$ in Fig. \ref{fig:cartoon}) as assumed in this work. Most likely it should have quasi-spherical shape and naturally asymmetries in the geometry of an astrophysical source are by definition sources of polarization. Furthermore, (as mentioned in Sect. \ref{subsec:results_time}) the shape and geometry of the event could in principle change as the event evolves. These are issues that are not addressed in this work. \citet{Patra2022} for example, suggest that the rise in the polarization of TDE AT~2019qiz within a month could potentially be explained by a change in the shape of the envelope. On the other hand, based on the modeling of their data, \citet{Leloudas2022} find that their studied TDEs show a convergence to an almost axisymmetric configuration soon after the flare peak. Of course the change in polarization could be a combination of geometrical and density/optical depth changes, making it challenging to disentangle which one has the biggest effect.

2) Throughout this work, we assume full-ionization within the outflow. The fact that our model is asymmetric with an emitting source offset from the center of the density distribution, could lead to an asymmetric electron density distribution at late times, when the heating source becomes weaker and the outflow becomes optically thin. That would make the electron scattering opacity not constant, but dependent on the density of free electrons which might be aspherical. In our modeling, we neglect this complication since the ionization fraction is very high (see discussion in Sect. \ref{sec:model}) and we assume that scattering remains the dominant source of opacity. Furthermore, the ionization fraction could decrease near the intersection point, since ionizing photons may not reach this high-density region. The drop of the ionization fraction in this region could potentially affect the polarization. However, as we have discussed in several parts of the manuscript, multiple scatterings (which is the case in the high-density regime around the intersection point) lead to a drop in polarization by randomizing the scattering angles. This would be the effect of thermalization as well, where photons would eventually be absorbed and re-emitted unpolarized.

3) Since we account for time evolution in this work, we should also account for changes in the emitting photosphere $\rph$. For simplicity, we neglected those changes to avoid having to introduce an extra parameter. Taking different ratios of $\rint/\rph$ might also probe the effect of a shrinking/expanding emitting photosphere. In addition, time-dependent effects are not considered and ionization is assumed to be constant throughout the propagation of the photons.

4) Throughout this work we assume a constant outflow velocity ($\vout=0.03c$) and, as discussed in Sect. \ref{sec:model}, this constant velocity used for simplicity here, does not take into
account the effect of the gravitational force from the black hole
and imposes that all the matter is not bound by the self-crossing
shock while more detailed calculations \citep{Lu2020} find that part of it stays bound. Furthermore, we haven't assumed a density profile for the outflow while the fallback rate rises to peak. However the physical processes that drive the light curve rise of TDEs are still debated hence there is not a standard model to use.

\section{Summary and Conclusions} \label{sec:conclusion}

In this work we model the continuum polarization levels of TDEs using the 3-D Monte Carlo radiative transfer code \textsc{possis}, based on the collision-induced outflow TDE emission scenario (CIO) where unbound shocked gas reprocesses the hard emission from the accretion flow into UV and optical bands. This work is timely as the first polarimetric observations of optical TDEs start to appear in the literature \citep{Leloudas2022,Patra2022} and it is necessary for any model trying to explain the TDE emission mechanism, to be able to reproduce the observed polarization values.

\begin{enumerate}[label={\arabic*.}]
\item We study two different cases of peak mass fallback rates $\mdotp$ at time $\tmin\sim40\rm d$ from the peak of the fallback; Case A with ${\mdotp = 2.93 \,\msun\, \rm yr^{-1}}$ and Case B with ${\mdotp = 0.29 \,\msun\, \rm yr^{-1}}$. A realistic mass rate and density profile for the outflow ($\mdotout$ and $\rho_{\rm out}$) launched from the intersection point leads to an envelope mass of $\menv\sim0.27\msun$ for Case A and $\menv\sim0.03\msun$ for Case B. We run twelve simulations for each case, for a realistic range of physical parameters $\rph$, $\rint$, $\delth$. 
\item For the twelve simulations of Case A, we find polarization below one per cent ($P<1\%$) for every viewing angle and for 10/12 simulations it is lower than 0.5\%. The absolute value of polarization reaches its maximum ($P_{\rm max}$) for equatorial viewing angles.
\item For the twelve simulations of Case B, the model can produce a wide range of polarization levels for different viewing angles and configurations. Four simulations predict values of $P>4\%$ with the maximum predicted value being $P\approx8.8\%$, while all the rest predict maximum values of at least $P=1\%$. The absolute value of polarization reaches its maximum ($P_{\rm max}$) for more polar viewing angles.
\item We find that the polarization depends strongly on i) the scattering optical depths at the central regions (between the emitting photosphere and the intersection point) set by the different $\mdotp$ values and ii) the viewing angle $\Theta$.
\item We study how the time evolution of the CIO model might affect the polarization by running simulations at time snapshots of $t=\tmin+25\rm d$ and $t=\tmin+50\rm d$. We find that as time passes, even though the mass in the envelope is increasing, the radius $\rmax$ expands and the densities and optical depths drop. This leads to higher maximum $P_{\rm max}$ values; for example the polarization reaches a striking $P\approx14\%$ for $\Theta\sim66^{\circ}$ at $t=\tmin+50$ days for a specific simulation setup. However the opposite trend can be observed for specific viewing angles.
\item We explore the dependence on the free geometrical parameters of our model ($\rint$, $\rph$, $\delth$) by varying only one of them. Extending the distance $\rint$ between the intersection point and the black hole seems to generally favour higher polarization levels. Widening the opening angle $\delth$ when $\rint/\rph\sim1$ seems to favour higher polarization levels as well. However if the distance between $\rph$ and $\rint$ becomes larger, widening the opening angle $\delth$ seems to favour lower polarization levels.
\item We estimate the $\mdotp$ values (based on observables) of TDEs AT~2018dyb \citep{Leloudas2022} and AT~2019qiz \citep{Patra2022}, in order to compare our model with actual observations. We find that the continuum polarization values of these TDEs (and its evolution with time) can indeed be reproduced by their matching $\mdotp$ cases probed in this work, Cases B for the former and Case A for the latter. Furthermore we attempt to constrain the viewing angles under which those TDEs are observed and we show that multi-epoch polarimetric observations can become a key factor in constraining the viewing angle of TDEs. 
\item The estimated $\mdotp$ values of AT~2019qiz match nicely with the values of Case A ($\mdotp = 2.93 \,\msun\, \rm yr^{-1}$).
Case A has been the case resulting to the lowest predicted $P$ values and AT~2019qiz has a $P\sim0$ at peak. On the other hand, the estimated $\mdotp$ values of AT~2018dyb match nicely with the values of Case B ($\mdotp = 0.29 \,\msun\, \rm yr^{-1}$). Case B has been the case resulting to the highest predicted $P$ values and AT~2018dyb has the highest $P$ up to date. These are encouraging findings for our model.   
\item Stars with $\mstar\geq1\,\msun$, should have $\mdotp\geq3\,\msun\, \rm yr^{-1}$ \citep{Law-Smith2020}. Based on our work, within the CIO scenario we expect that tidal disruptions of stars of $\mstar\geq1\,\msun$, should have continuum polarization levels below $P=1\%$ and for most of the cases below $P=0.5\%$, at least for times around $t=\tmin$. 
\end{enumerate}

Degeneracies between the mass outflow rates and the potential viewing angles could be broken if we could put better constraints on the $\mdotp$ of observed TDEs. Combined with other observables such as X-ray emission properties and spectroscopic features \citep{Dai2018,Charalampopoulos2022}, polarimetric observations and comparison with modeling results (like the ones that this work provides) could help us to put firm constraints on the viewing angle that TDEs are observed from. Naturally, more polarimetric observations of TDEs are encouraged in order to better constrain the parameters of our model and, as we showed, multi-epoch observations are very critical for better constraining the viewing angle and break degeneracies.

\begin{acknowledgements}

We thank the anonymous referee for comments that helped improve this paper. P.C and G.L are supported by a research grant (19054) from VILLUM FONDEN. M. B. acknowledges support from the Swedish Research Council (Reg. no. 2020-03330) and from the European Union’s Horizon 2020 Programme under the AHEAD2020 project (grant agreement n. 871158). This project has received funding from the European Union’s Horizon 2020 Framework Programme under the Marie Sklodowska-Curie grant agreement no. 836751.

\end{acknowledgements}

\bibliographystyle{aa}
\bibliography{bib.bib}

\appendix{}
\onecolumn
\centering

\section{Graphs} \label{apdx:graphs}
\begin{figure*}[h]
\centering
\includegraphics[width=0.84 \textwidth]{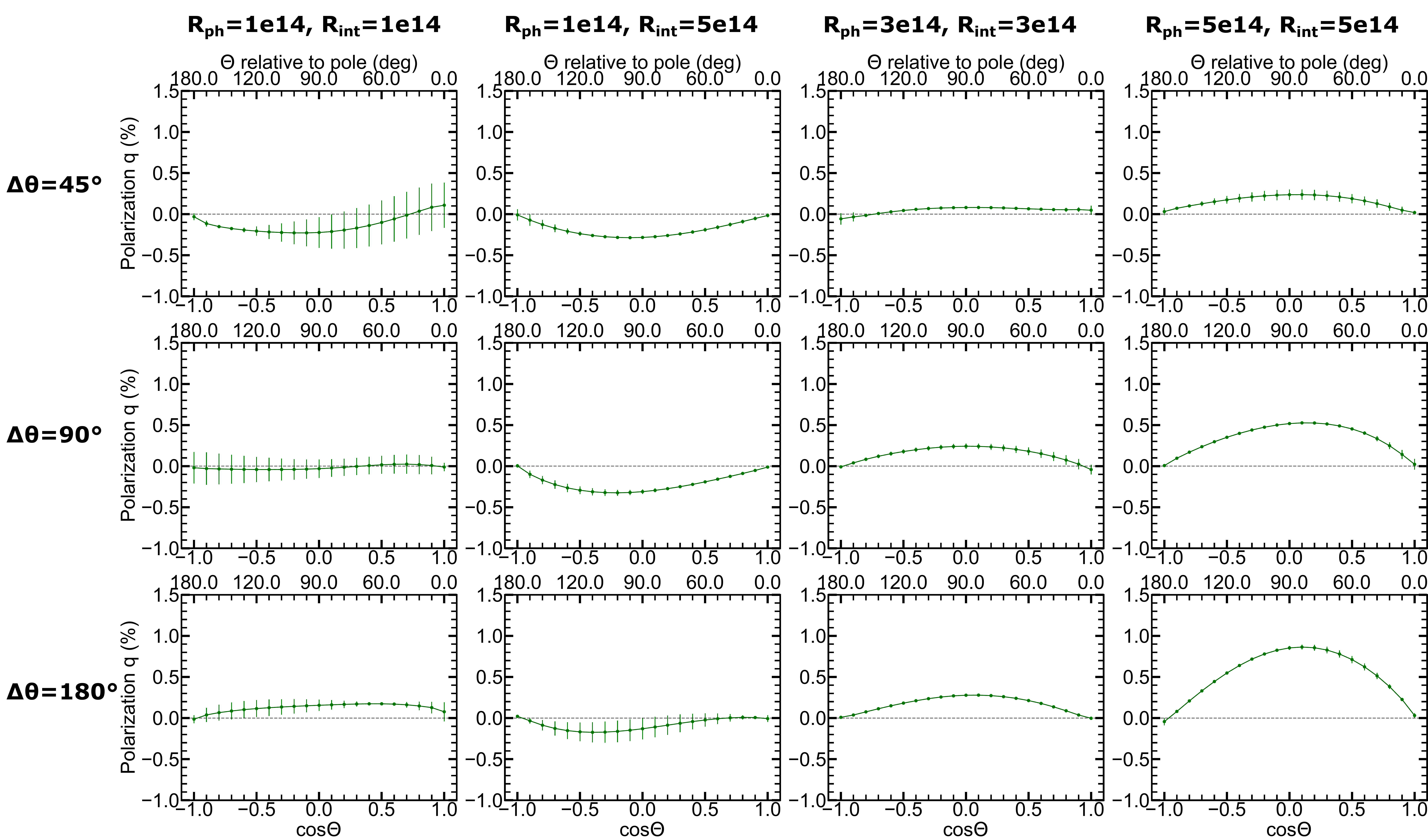}
\caption{Same as Fig. \ref{fig:med_mass_subs} (Case A, high mass case) but we change the first column to $\rph=10^{14}\, {\rm cm}$, $\rint=10^{14}\, {\rm cm}$ and the third column to $\rph=3\times10^{14}\, {\rm cm}$, $\rint=3\times10^{14}\, {\rm cm}$. We keep the second and forth column the same as in the original Figure in order to facilitate visual comparison.}

\label{fig:med_mass_all_new}
\end{figure*}

\begin{figure*}
\centering
\includegraphics[width=0.84 \textwidth]{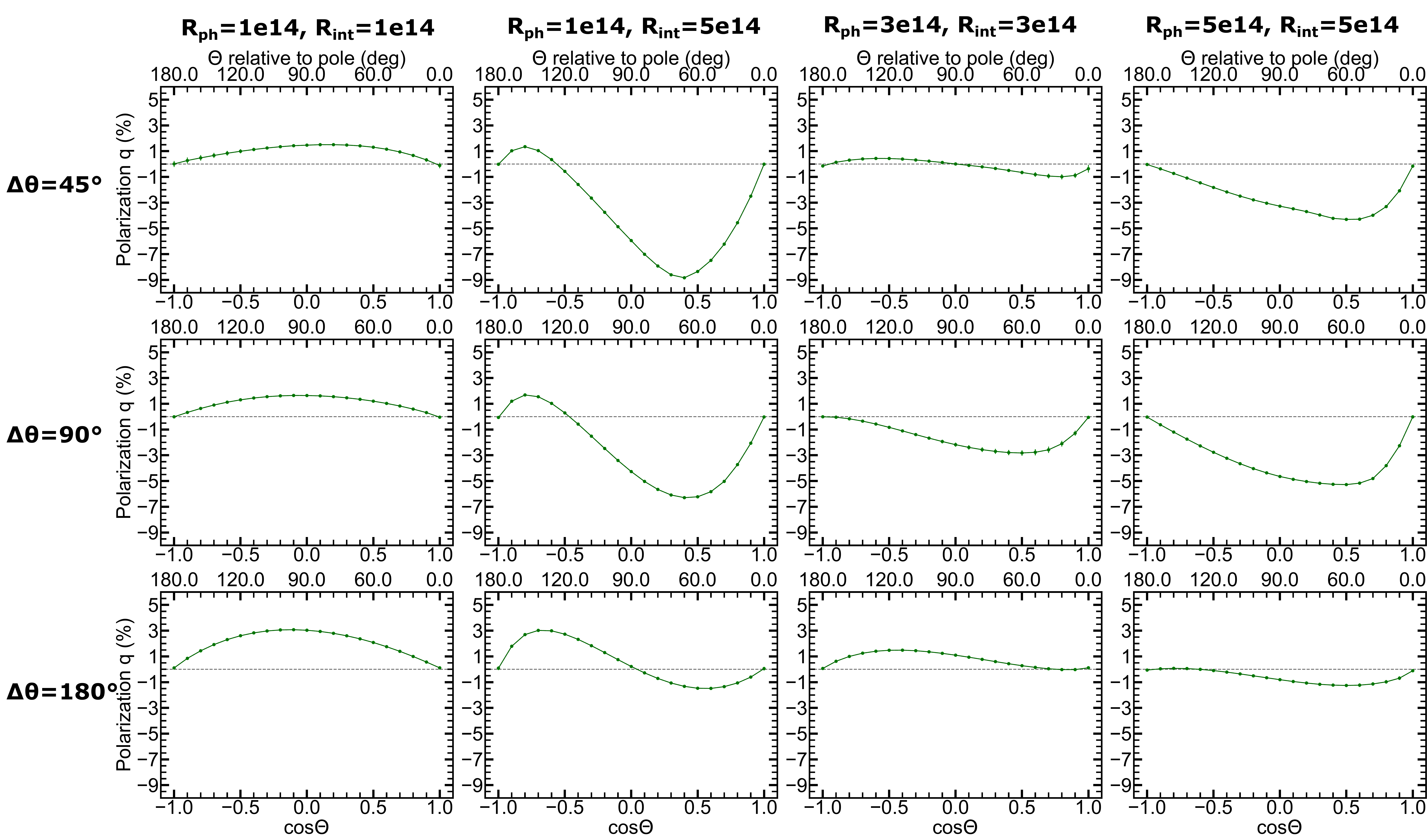}
\caption{Same as Fig. \ref{fig:low_mass_subs} (Case B, $\mdotp = 0.29 \,\msun\, \rm yr^{-1}$) but we change the first column to $\rph=10^{14}\, {\rm cm}$, $\rint=10^{14}\, {\rm cm}$ and the third column to $\rph=3\times10^{14}\, {\rm cm}$, $\rint=3\times10^{14}\, {\rm cm}$. We keep the second and forth column the same as in the original Figure in order to facilitate visual comparison.}
\label{fig:low_mass_all_new}
\end{figure*}

\begin{figure}
\centering
\includegraphics[width=0.6 \textwidth]{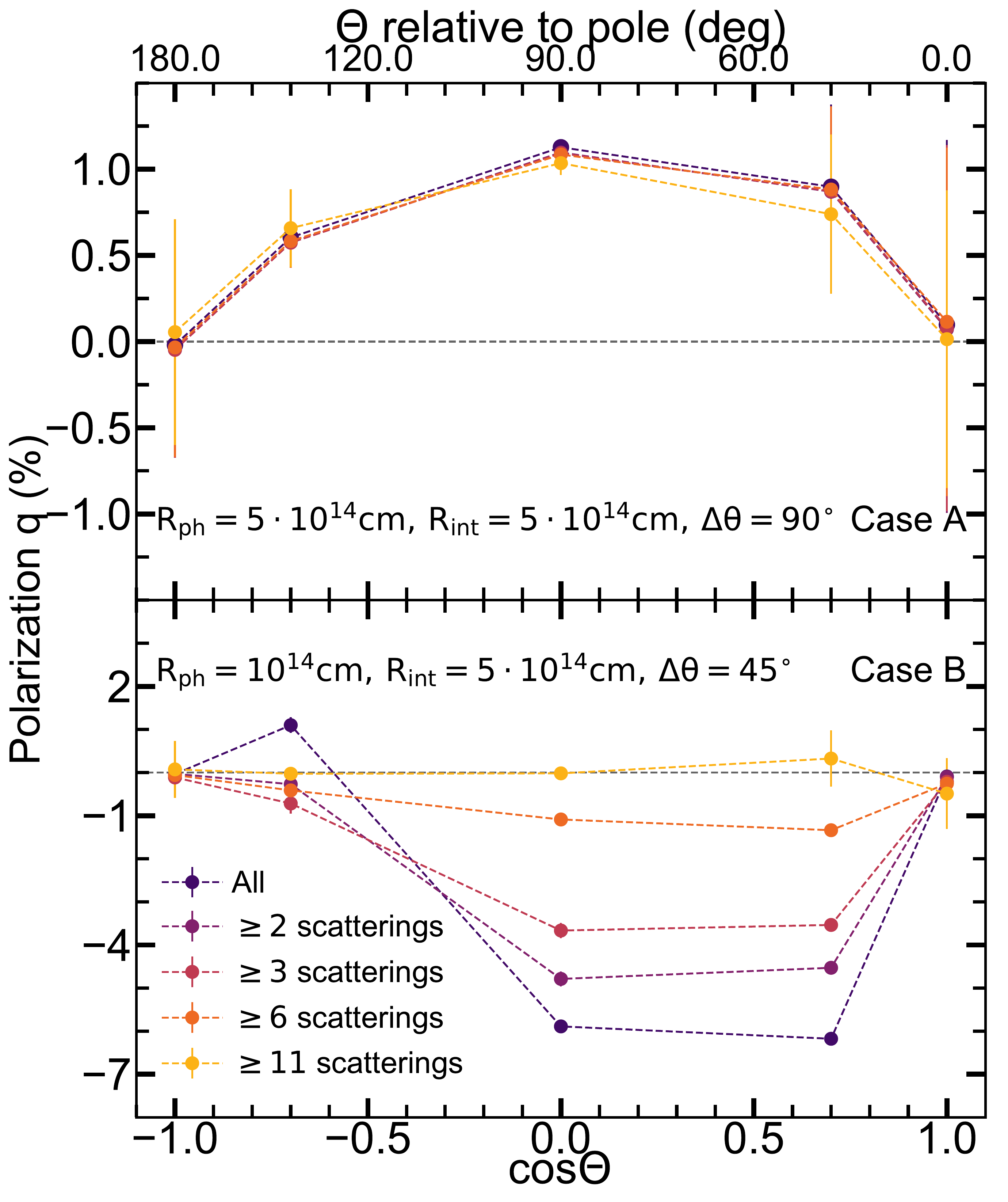}
\vspace{0cm}
\caption{Polarization as a function of the viewing angle $\Theta$
for five different viewing angles: $\Theta=180^{\circ},135^{\circ},90^{\circ},45^{\circ}\, {\rm and}\, 0^{\circ}$. In each different colored curve, we remove the contribution of photons that scattered less times than what is written in the legend. Top panel: Simulation s2 from Fig. \ref{fig:med_mass_explain}. Removing photons that scatter 10 or less times minimally affects the result. The polarization level is dominated by the multiple-scattered photons. Bottom panel: Same as top panel but for the simulation s3 of Fig. \ref{fig:low_mass_explain}. In the lower density regime of Case B, the polarization is dominated by emerging photons that experienced $\lesssim5$ scatterings. The positive $q$ values around the viewing angles of $\Theta=135^{\circ}$ are set by photons that scattered only once.}
\label{fig:scat}
\end{figure}


\begin{figure}
\centering
\includegraphics[width=0.49 \textwidth]{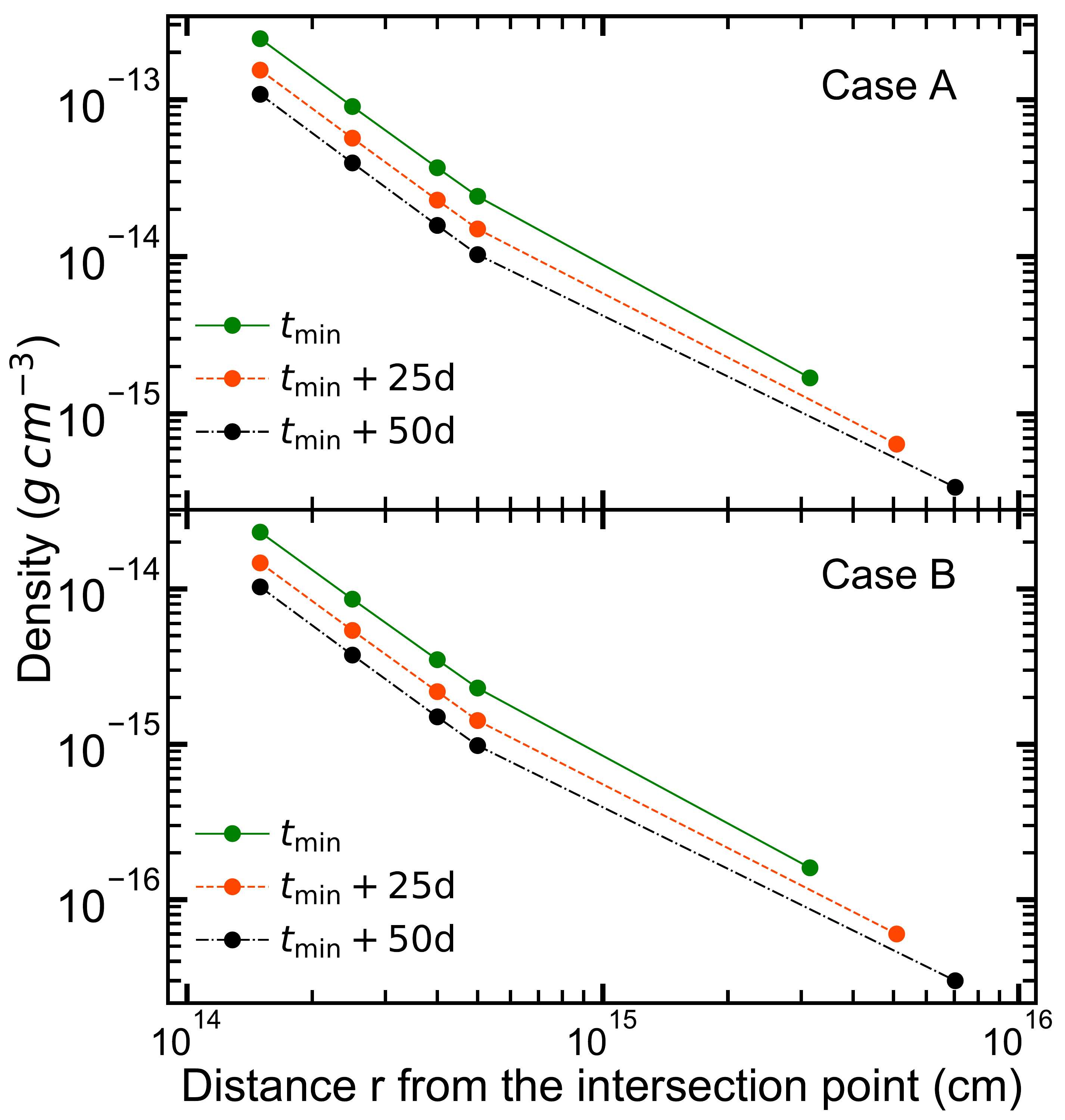}
\vspace{0cm}
\caption{Indicative densities at various distances $r$ from the intersection point, for the two different mass outflow rate cases (Case A top panel and Case B bottom panel) and for three different time snapshots ($t=\tmin$, $t=\tmin+25\rm d$ and $t=\tmin+50\rm d$). The outermost right point of every curve is where the $\rmax$ is located, that is, the outer radius of the outflow/grid. The exact numbers shown in this plot can be found in Table \ref{tab:densandtau} in the Appendix.}
\label{fig:dens_vs_r}
\end{figure}

\begin{figure*}
        \centering
        \begin{subfigure}[b]{1\textwidth}
            \centering
        \includegraphics[width=0.492 \textwidth]{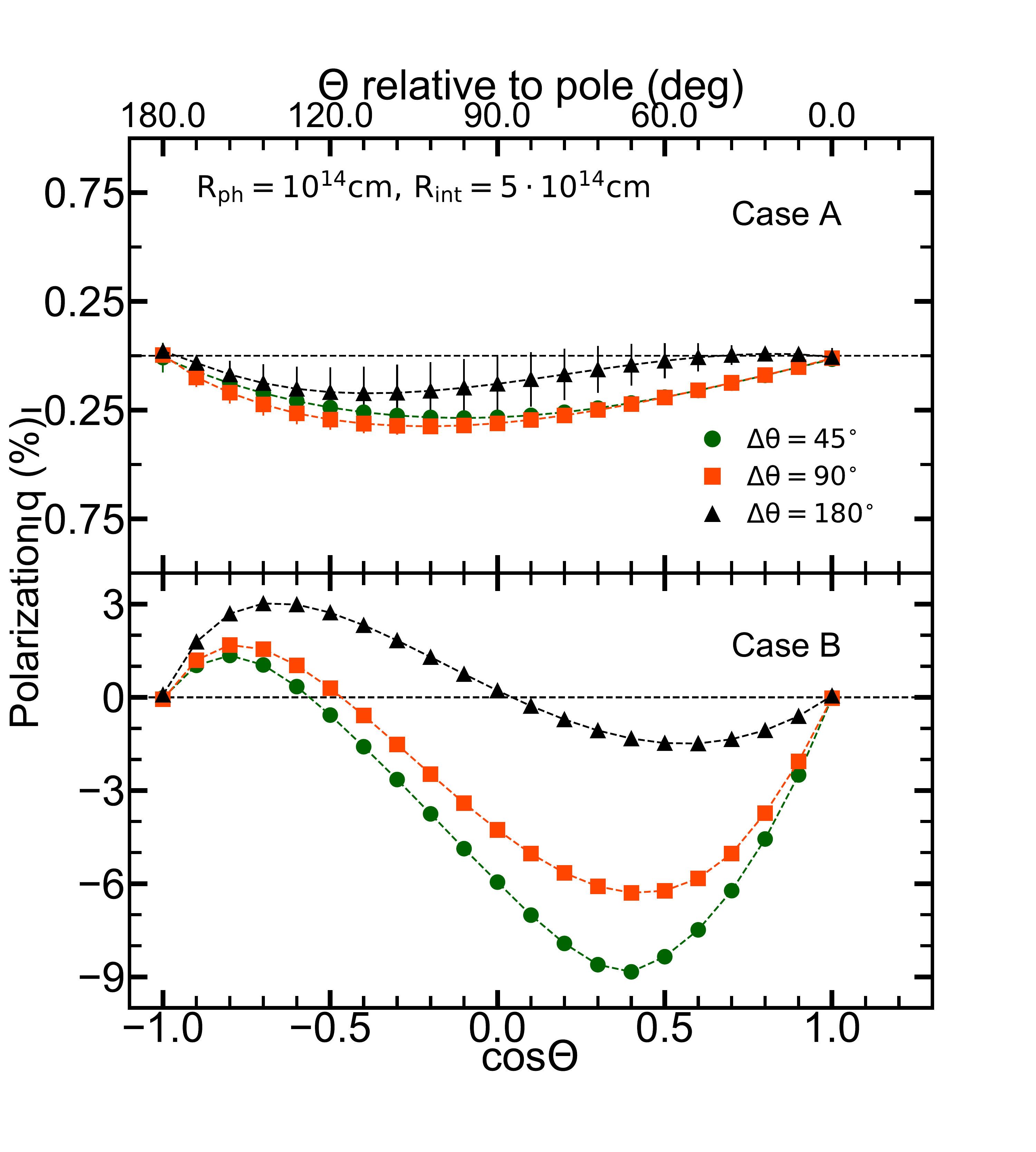}
        \includegraphics[width=0.492 \textwidth]{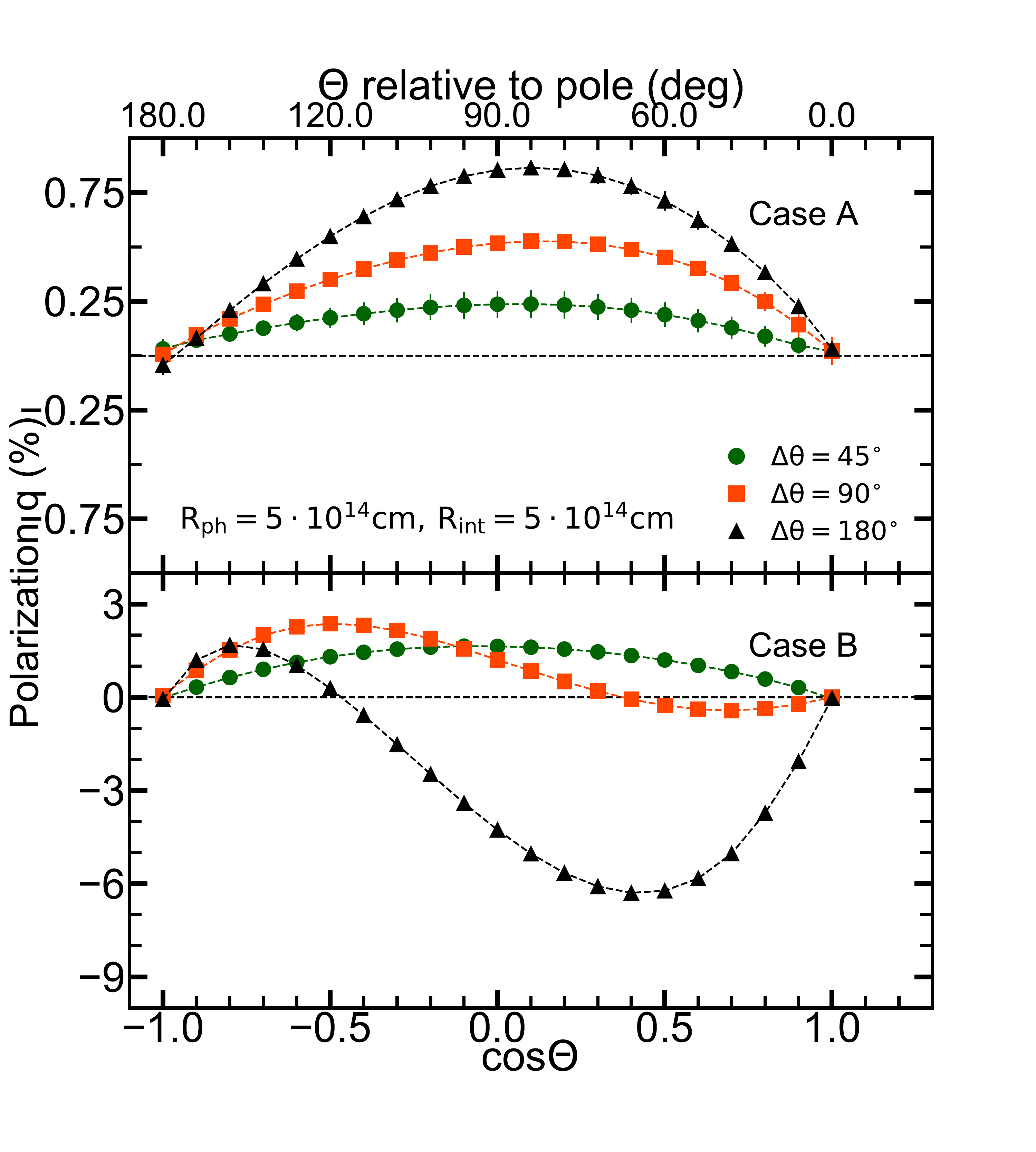}
        \end{subfigure}
        \caption{Investigating the polarization as a function of the viewing angle $\Theta$ for varying $\delth$ while we keep $\rph$ and $\rint$ fixed (Left panel: $\rphs$, $\rintl$, Right panel: $\rphl$, $\rintl$). The upper panels are for the high mass Case A and the lower panels for the low mass Case B. Different distances between $\rph$ and $\rint$ affect the outcome of varying $\delth$.}
        \label{fig:delths}
    \end{figure*}

\onecolumn
\centering

\section{Tables}\label{apdx:tables}

\begin{table}[h]
\renewcommand{\arraystretch}{1.4}
\setlength\tabcolsep{0.3cm}
\fontsize{10}{11}\selectfont
\caption{Indicative densities and optical depths located at various distances $r$ from the intersection point for the two different mass outflow rate cases and for three different time snapshots; $t=\tmin$ (left), $t=\tmin+25\rm d$ (middle) and $t=\tmin+50\rm d$ (right). }\label{tab:densandtau}
\begin{tabular}{l | c c | c c | c c }
\hline
 & ${\rho_{\rm out}}$ & $\tau = \int_{r}^{\infty} \rho\kappa_{es} \,dr$ & ${\rho_{\rm out}}$ & $\tau = \int_{r}^{\infty} \rho\kappa_{es} \,dr$ & ${\rho_{\rm out}}$ & $\tau = \int_{r}^{\infty} \rho\kappa_{es} \,dr$  \\
 & ($10^{-15}\, gcm^{-3}$ ) & & ($10^{-15}\, gcm^{-3}$ ) && ($10^{-15}\, gcm^{-3}$ )  \\
\noalign{\global\arrayrulewidth=0.7mm}\hline
\noalign{\global\arrayrulewidth=0.4pt}
\hline
\bf Case A & \multicolumn{2}{c|}{$\bf t=t_{\rm \bf min}$}&  \multicolumn{2}{c|}{$\bf t=t_{\rm \bf min}+25d$}  & \multicolumn{2}{c}{$\bf t=t_{\rm \bf min}+50d$}\\
\hline
$\rm r=1.5\times10^{14}\, {\rm cm}$&244 &15.72 &154 &9.49&108 &6.42 \\
$\rm r=2.5\times10^{14}\, {\rm cm}$&90.3 &10.81 &56.8 &6.4&39.5 &4.26 \\
$\rm r=4\times10^{14}\, {\rm cm}$&36.8 &7.96 &22.9 &4.61&15.8 &3.02 \\
$\rm r=5\times10^{14}\, {\rm cm}$&24.2 &6.97 &15 &3.99&10.3 &2.6  \\
$\rm r=\dsc$&- &1 &- &1&0.47 &1\\
$\rm r=\rmax$&1.69 &2.76&0.64 &1.15 &0.34 &0.57\\
\noalign{\global\arrayrulewidth=0.7mm}\hline
\noalign{\global\arrayrulewidth=0.4pt}
\hline
\bf Case B & \multicolumn{2}{c|}{$\bf t=t_{\rm \bf min}$}&  \multicolumn{2}{c|}{$\bf t=t_{\rm \bf min}+25d$}  & \multicolumn{2}{c}{$\bf t=t_{\rm \bf min}+50d$} \\
\hline
$\rm r=1.5\times10^{14}\, {\rm cm}$&23.2 &1.5 &14.7 &0.9&10.3 &0.61 \\
$\rm r=2.5\times10^{14}\, {\rm cm}$&8.59 &1.03 &5.4 &0.61&3.75 &0.41 \\
$\rm r=4\times10^{14}\, {\rm cm}$&3.5 &0.76 &2.18 &0.44&1.5 &0.29 \\
$\rm r=5\times10^{14}\, {\rm cm}$&2.3 &0.66 &1.42 &0.38&0.98 &0.25 \\
$\rm r=\dsc$&7.96 &1 &19.5 &1 &28.2 &1\\
$\rm r=\rmax$&0.16 &0.26&0.06 &0.11 &0.03 &0.05\\
\noalign{\global\arrayrulewidth=0.7mm}\hline
\noalign{\global\arrayrulewidth=0.4pt}
\end{tabular}
\end{table}

\end{document}